\documentclass[aip,jcp,reprint,amsmath,amssymb]{revtex4-1}
\usepackage{graphicx}
\usepackage{multirow}

\usepackage[normalem]{ulem}

\usepackage{xcolor}

\definecolor{vbred}{rgb}{1.0, 0.0, 0.0}

\begin{document}


\title{Efficient All-electron Hybrid Density Functionals for Atomistic Simulations Beyond 10,000 Atoms}

\author{Sebastian Kokott}
\affiliation{The NOMAD Laboratory at the Fritz Haber Institute of the Max-Planck-Gesellschaft and IRIS Adlershof of the Humboldt-Universit\"{a}t zu Berlin, Germany}

\author{Florian Merz}
\affiliation{Lenovo HPC Innovation Center, Stuttgart, Germany}

\author{Yi Yao}
\affiliation{Thomas Lord Department of Mechanical Engineering and Material Science, Duke University, Durham, North Carolina 27708, USA}

\author{Christian Carbogno}
\affiliation{The NOMAD Laboratory at the Fritz Haber Institute of the Max-Planck-Gesellschaft and IRIS Adlershof of the Humboldt-Universit\"{a}t zu Berlin, Germany}

\author{Mariana Rossi}
\affiliation{MPI for the Structure and Dynamics of Matter, Luruper Chaussee 149, 22761 Hamburg, Germany}

\author{Ville Havu}
\affiliation{Department of Applied Physics, School of Science, Aalto University, Espoo, Finland}

\author{Markus Rampp}
\affiliation{Max Planck Computing and Data Facility, 85748 Garching, Germany}

\author{Matthias Scheffler}
\affiliation{The NOMAD Laboratory at the Fritz Haber Institute of the Max-Planck-Gesellschaft and IRIS Adlershof of the Humboldt-Universit\"{a}t zu Berlin, Germany}

\author{Volker Blum}
\affiliation{Thomas Lord Department of Mechanical Engineering and Material Science, Duke University, Durham, North Carolina 27708, USA}
\affiliation{Department of Chemistry, Duke University, Durham, North Carolina 27708, USA}

\date{\today}

\begin{abstract}
Hybrid density functional approximations (DFAs) offer compelling accuracy for {\it ab initio} electronic-structure simulations of molecules, nanosystems, and bulk materials, addressing some deficiencies of computationally cheaper, frequently used semilocal DFAs.
However, the computational bottleneck of hybrid DFAs is the evaluation of the non-local exact exchange contribution, which is the limiting factor for the application of the method for large-scale simulations. 
In this work, we present a drastically optimized resolution-of-identity-based real-space implementation of the exact exchange evaluation for both non-periodic and periodic boundary conditions in the all-electron code FHI-aims, targeting high-performance CPU compute clusters. The introduction of several new refined Message Passing Interface (MPI) parallelization layers and shared memory arrays according to the MPI-3 standard were the key components of the optimization. We demonstrate significant improvements of memory and performance efficiency, scalability, and workload distribution, extending the reach of hybrid DFAs to simulation sizes beyond ten thousand atoms. As a necessary byproduct of this work, other code parts in FHI-aims have been optimized as well, e.g., the computation of the Hartree potential and the evaluation of the force and stress components. We benchmark the performance and scaling of the hybrid DFA based simulations for a broad range of chemical systems, including hybrid organic-inorganic perovskites, organic crystals and ice crystals with up to 30,576 atoms (101,920 electrons described by 244,608 basis functions).
\end{abstract}

\maketitle 


\begin{figure*}[t!]
    \centering
    \includegraphics[width=0.9\textwidth]{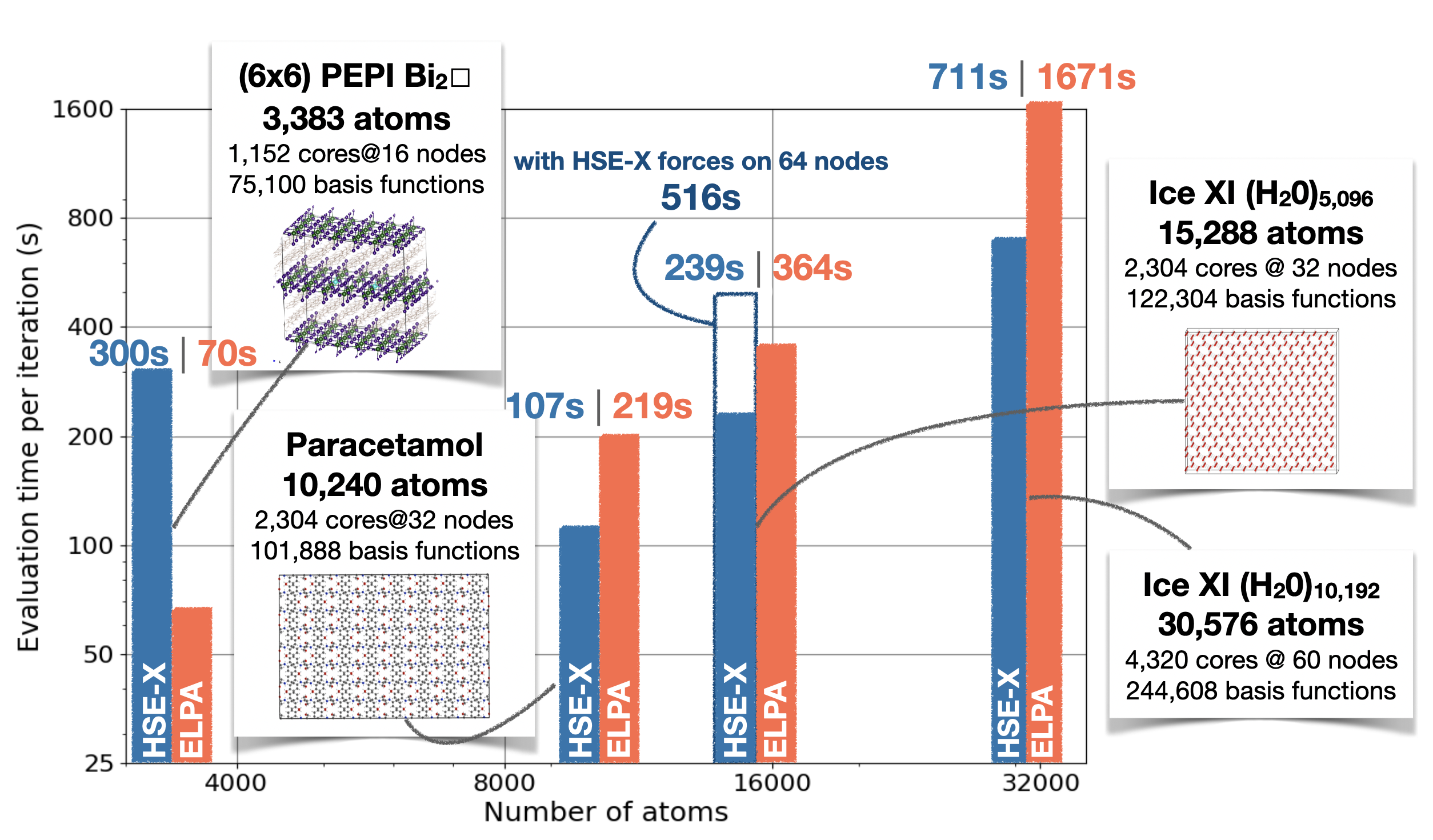}
    \caption{Benchmark results for the largest periodic structures considered in this work. Average runtimes to evaluate the HSE06 exchange operator (blue bars) and the ELPA two stage eigenvalue solver (red bars) per self-consistent field iteration are shown. The HSE06 hybrid functional was used for all simulations. The following systems were simulated (from left to right): phenylethylammonium lead iodide (PEPI) with a defect complex (as indicated by the square in the chemical formula)~\cite{Lu2023}, a $4\times4\times4$ paracetamol supercell, a 15,288-atoms Ice XI supercell (including a force evaluation), and a 30,576-atom Ice XI supercell. All calculations were carried out on the Raven HPC cluster at the MPCDF using Intel Xeon IceLake (Platinum 8360Y) nodes with 72 cores per node.}
    \label{fig:large-scale-overview}
\end{figure*}

\section{Introduction}
\label{SecIntro}
Density-functional theory (DFT) and its approximations (DFAs) have shaped the fields of computational chemistry and materials science by providing a powerful framework to investigate molecules, nanosystems, solids, and surfaces at the atomic scale. The scaling for finding the solution to the Kohn-Sham equations for (semi)local DFAs is formally $O(N^3)$, where $N$ is a measure of the system size, when using direct eigensolvers. In practice, the actual scaling can often be reduced to $O(N^2)$ when the cache memory of modern CPUs is efficiently utilized, as,~e.g.,~demonstrated by the ELPA eigensolver~\cite{Yu2021,Kus2019,Marek.2014,Marek2014}. Recent developments for linear scaling DFT have driven the field to system sizes of dizzying orders of magnitudes (up to many tens of millions of atoms).\cite{Schade2023,nakata2020large} The key to achieving linear-scaling in electronic structure methods is exploiting locality, since localized basis functions with finite spatial extent lead to sparsity in the density matrix. Some prominent choices for localized basis functions are numeric atom-centered orbitals (NAOs)~\cite{luo2020parallel}, non-orthogonal generalized Wannier functions~\cite{prentice2020onetep}, polarized atomic orbitals~\cite{nakata2020large}, and Gaussian functions. For all these approaches, large-scale, semilocal DFT calculations with linear scaling were successfully demonstrated.

Local and semilocal DFAs, such as the local density approximation (LDA), generalized gradient approximations (GGAs), and meta-GGAs, often face accuracy limitations in predictions of important chemical and physical properties, especially when charge transfer or localisation play an important role~\cite{goerigk2010,Tran2020}. To overcome these challenges and enhance the predictive capabilities of DFT, hybrid DFAs\cite{Becke1993,ernzerhof1999,heyd2003} have long been employed. For many systems, hybrid DFAs significantly improve the prediction of electronic properties, e.g., band gaps~\cite{garza2016}, charge localization~\cite{Lany2010,Kokott2018}, or the description of d-orbitals~\cite{Finazzi2008}. The key ingredient to hybrid DFAs is mixing the (semi)local exchange of LDA, GGAs, or meta-GGAs with some fraction of non-local exact exchange (EXX). Additional flexibility is provided by treating only a certain spatial range of the Coulomb operator non-locally within the framework of the hybrid density functionals, while keeping the remainder semilocal: By introducing a range-separation function for the Coulomb potential, a variety of different functionals can be constructed, e.g., HSE06~\cite{heyd2003,heyd2006,Krukau2006}, LC-wPBEh~\cite{vydrov2006assessment}, M11~\cite{peverati2011}, wB97~\cite{Chai2008}. Because of its smoothness, the error function is a frequent choice to divide the Coulomb potential into long- and short-range parts:
\begin{align}
    v(r) = \underbrace{\frac{1-\text{erf}(\omega r)}{r}}_{v_\text{SR}(r)} + \underbrace{\frac{\text{erf}(\omega r)}{r}}_{v_\text{LR}(r)} \, .
    \label{eq:coulomb_kernel}
\end{align}
Here, $r = \vert\mathbf{r}-\mathbf{r}^{\prime}\vert$, $\omega$ (an adjustable inverse length) is the range-separation parameter, and $v_\text{SR}(r)$ and $v_\text{LR}(r)$ are the short- and long-range Coulomb potential, respectively. Other physics-inspired range separation strategies are possible as well~\cite{lorke2020koopmans}. Below, we will refer to the individual range-separated parts of the Coulomb potential in the exchange operator as Coulomb kernels. In general, we can denote the fractions of non-local full exchange and of nonlocal short-range exchange by two parameters $\alpha$ and $\beta$, respectively. Thus, the following contributions to the exchange energy $E_x$ are obtained:
\begin{align}
    E_x(\alpha,\beta,\omega) & =  \alpha E_\text{EXX} + \beta E^\text{SR}_\text{EXX}(\omega) \nonumber \\
    & + (1-\alpha) E_\text{x-DFA} - \beta E^\text{SR}_\text{x-DFA}(\omega) \, .
\end{align}
 $E_\text{EXX}$ is the EXX energy using the full Coulomb potential and $E_\text{EXX}(\omega)$ is the short-range EXX energy. Similarly, $E_\text{x-DFA}$ is the semilocal DFA exchange energy for the full-range Coulomb operator and $E^\text{SR}_\text{x-DFA}(\omega)$ is the short-range semilocal DFA exchange energy.
Using this notation, the PBE0 functional~\cite{ernzerhof1999,adamo1999} can be recovered by choosing $\alpha=0.25$ and $\beta=0$ and a typical version of the HSE06 functional~\cite{heyd2003,heyd2006} benchmarked by Kurkau {\it et al.}~\cite{Krukau2006} can be obtained by setting $\alpha=0.0$, $\beta=0.25$, and $\omega=0.11$ Bohr$^{-1}$. The long-range corrected LC-wPBEh~\cite{vydrov2006assessment} and the long-range corrected B97 functional wB97~\cite{Chai2008} require $\alpha=1.0$, $\beta=-1.0$, and $\omega=0.4$ and choosing PBE or B97 as GGA functionals, respectively. In order to cover families of functionals with $\alpha \neq 0$ and $\beta \neq 0$, it can be convenient to compute two EXX matrices within a single call to a first-principles code -- one matrix for each Coulomb kernel. In the following we refer to all types of screened (long- and short-range) and un-screened (full) EXX contributions simply as EXX contributions. The difference between these different types of EXX contributions lies just in the shape of the screened or unscreened Coulomb potential (i.e., the Coulomb kernel) and the same algorithm can be employed to evaluate the exchange contribution.

Along with the increase in accuracy, hybrid DFAs typically result in significantly larger computational cost compared to semilocal DFAs due to the need to evaluate the non-local exchange operator. In fact, a na\"ive implementation of the electron-repulsion integrals formally scales with $O(N^4)$ with system size $N$. To overcome this hurdle and enable linear-scaling ($O((N)$) hybrid DFT calculations for extended systems, various strategies have been successfully employed, e.g., linear scaling incremental Fock builds~\cite{Schwegler1997}, the Linear exchange K (LinK) approach~\cite{Ochsenfeld1998}, resolution-of-identity schemes (e.g., Refs.~\onlinecite{ren2012}, \onlinecite{ihrig2015}, \onlinecite{Bussy2024}, \onlinecite{forster2020double} and references therein), auxiliary density matrix methods~\cite{hu2017interpolative}, non-orthogonal generalised Wannier functions~\cite{Dziedzic2013}, transformations to maximally localized Wannier functions~\cite{wu2009order,ko2020enabling}, and  adaptive compression in a low-rank decomposition~\cite{lin2016adaptively}. However, the computational and book-keeping overhead that incurs in such linear-scaling approaches leads to considerably higher prefactors and more complex code, typically hindering an efficient parallelization in terms of memory and computation time. Accordingly, hybrid DFT calculations are still typically considerably more costly than standard semilocal DFAs. In Table~\ref{tab:records}, we summarize some literature examples of large-scale hybrid DFT calculations, including the codes and methods that were employed. Evidently, several codes and implementations can facilitate hybrid DFT calculations up to several thousands of atoms in size on modern HPC architectures.

\begin{table*}[]
    \centering
    \begin{tabular}{l|l|l|l}
        Code name & System (Number of atoms) & Method & Reference \\
        \hline \hline
        FHI-aims & (H$_2$O)$_{10,192}$ (30,576 atoms) & NAOs with localized resolution of identity & This work. \\
        \hline
        Quantum Espresso & (H$_2$O)$_{512}$ (1,536 atoms) & MLWF with adaptively compressed exchange & Ref.~\onlinecite{ko2023high} \\
         CP2K & Rubredoxin (2,825 atoms) & GPW and auxiliary density matrix methods & Ref.~\onlinecite{guidon2010auxiliary} \\
         ONETEP & Stacked polymer chains (2,000 atoms) & NGWFs and spherical waves resolution of identity & Ref.~\onlinecite{prentice2020} \\
         BigDFT & (H$_2$O)$_{512}$ (1,536 atoms) & Wavelets with GPU acceleration & Ref.~\onlinecite{ratcliff2018affordable} \\
         CRYSTAL & Amorphous silica MCM-41 (4,632 atoms) & GTOs & Ref. \onlinecite{erba2017large} \\ 
    \end{tabular}
    \caption{Some published large-scale hybrid DFT calculations for different methods and codes at the time of writing. The selection is restricted to simulations with three dimensional periodic boundary conditions. NAOs: numeric atom-centered orbitals, MLWF: maximal localized Wannier functions, NGWFs: Non-orthogonal Generalized Wannier Functions, GTOs: Gaussian-type orbitals.}
    \label{tab:records}
\end{table*}

This work describes recent algorithmic improvements achieved for the EXX contributions that drastically accelerate hybrid DFT calculations for large systems (non-periodic and periodic) on existing massively parallel CPU clusters, without introducing any new approximations. The approach is implemented in the all-electron code FHI-aims~\cite{blum2009,ren2021all,ihrig2015,levchenko2015,gavini2023roadmap} using numeric atom-centered orbitals (NAOs) as basis functions, but the underlying techniques are general and amenable to any other code using localized orbitals for discretization. Specifically, we build on the localized resolution-of-identity (also sometimes referred to as density fitting) implementation originally described as RI-LVL in Refs.~\onlinecite{ihrig2015,levchenko2015} and referred to as ``2015 implementation'' below. Exploiting localization is key to achieving high performance and a low memory footprint in the evaluation of the EXX contribution. In the limit of large system sizes and for periodic systems with a band gap, the long-range tail of the Coulomb potential Eq.~\eqref{eq:coulomb_kernel} will be suppressed in the exchange term because of the finite range of the density matrix~\cite{benzi2013decay}. Thus, the EXX term becomes effectively localized. In conjunction with an appropriate choice of localized basis functions, the EXX matrix Eq.~\eqref{eq:realspace_exchange} becomes sparse in real space and can be evaluated at a computational cost that scales linearly with system size. 
In addition, we describe further algorithmic improvements in the code regarding the evaluation of the Hartree potential, the evaluation of the Pulay force terms, and the initialization of general index arrays for periodic boundary conditions, which could otherwise become bottlenecks at certain regimes with the new hybrid-functional implementation. 

For most of the paper, we will focus on the HSE06 functional, which only uses the short-range Coulomb potential $v_\text{SR}(r)$. This is a very popular functional that provides a good balance between accuracy and computational performance (time and memory) in large-scale simulations, due to the restriction of EXX exchange contributions to a smaller range. For comparison, we also show the performance of the global hybrid functional PBE0. In all cases, the solution of the generalized Kohn-Sham equations is obtained with the direct eigensolver ELPA~\cite{Yu2021,Kus2019,Marek.2014,Marek2014}, version 2023.05.001. Some key examples of system types and sizes that are now attainable are visualized in Figure~\ref{fig:large-scale-overview}, ranging up to 30,576 atoms in size. Details of these and further benchmarks are provided in Sec.~\ref{SecBenchmark} below.

The impact of our work for physics applications will be significant, since simulations of very large, complex systems using hybrid DFAs are now affordable on typical high-performance computing resources. In cases where hybrid DFAs matter, e.g., for energy level alignments in complex structures~\cite{Liu2018,park2023thickness}, the added accuracy of hybrid DFAs can be essential. One example which made use of the hybrid DFT improvements described here is a recent study addressing isolated substitutional defects and defect complexes in a layered hybrid perovskite crystal, phenethylammonium lead iodide~\cite{Lu2023} (PEPI, also included in Figure~\ref{fig:large-scale-overview}). In order to eliminate any relevant interactions of defects across supercell boundaries, structure sizes up to 3,383 atoms were employed, providing direct access to the spin-orbit coupled DFT-HSE06 energy band structure and associated defect energy levels. In contrast, smaller supercell models were shown to be insufficiently large, even when many hundred atoms were included, since clear dispersion features of the defect states demonstrated the presence of noticeable defect-defect interactions across unit cell boundaries. Affordable simulations of systems spanning thousands of atoms using hybrid DFAs will be equally beneficial in many other scenarios where the environment of a localized defect or chemical process needs to be sufficiently large to enable realistic results, particularly when energy levels are at issue. Our development paves the way for such simulations across chemistry and materials science.

The paper is structured as follows: First, we introduce the formulas needed for the evaluation of the EXX operator. Based on them, we describe the algorithm and the improvements that have been made compared to the earlier RI-LVL EXX implementation in Ref.~\onlinecite{levchenko2015}. Then, the strong and weak scaling behaviors of the new implementation are discussed. Finally, we show benchmarks of the improved implementation for a broad range of systems covering solids, surfaces, nanosystems, clusters, and molecules.

\section{Description of the real-space formalism}

We here briefly outline the notation and formalism of the real-space evaluation of the EXX operator as implemented in FHI-aims. The basic equations are those of the initial linear-scaling implementation of Levchenko {\it et al.}~\cite{levchenko2015}. Thus, we use the notation introduced in that reference and only briefly summarize the key expressions and refer to Ref.~\onlinecite{levchenko2015} for details. The formalism works for both periodic and non-periodic systems. In the following, we present the more general formulae that account for the periodic case. The non-periodic case can be recovered by considering only $\mathbf{R} = 0$, i.e., by omitting any $\mathbf{k}$ points and Bloch sums over unit cells.

In generalized Kohn-Sham theory, the $\mathbf{k}$-dependent EXX operator $K$ or a fraction thereof is added to the Hamiltonian. Elements of the $K$ operator are given by
\begin{align}
    K^{\sigma}_{ij} (\mathbf{k}) = \sum_\mathbf{R} e^{i\mathbf{k}\cdot\mathbf{R}} X^{\sigma}_{ij}(\mathbf{R}),
    \label{eq:fock_operator_k}
\end{align}
where the Latin symbols $i$, $j$ denote the NAO basis functions and $\sigma$ the spin index. The vector $\mathbf{k}$ refers to a point of the $\Gamma$-centred k-grid and $\mathbf{R}$ is a  real-space lattice vector. Using the localized resolution-of-identity (RI) approach, called RI-LVL~\cite{ihrig2015}, the exchange operator in real-space $X_{ij}(\mathbf{R})$ can be written as follows:
\begin{align}
X^{\sigma}_{i j}(\mathbf{R})=& \sum_{k \mathbf{R}^{\prime}} \sum_{\mathbf{R}^{\prime \prime}} \sum_{\mu \mathbf{Q}^{\prime}} \sum_{\nu \mathbf{Q}^{\prime \prime}} C_{i k\left(\mathbf{R}^{\prime}\right)}^{\mu(\mathbf{Q}^{\prime})} V_{\mu \nu\left(\mathbf{R}+\mathbf{Q}^{\prime \prime}-\mathbf{Q}^{\prime}\right)} C_{j l\left(\mathbf{R}^{\prime \prime}\right)}^{\nu(\mathbf{Q}^{\prime \prime})} \nonumber \\
&\times D_{k l}^{\sigma}\left(\mathbf{R}+\mathbf{R}^{\prime \prime}-\mathbf{R}^{\prime}\right)
\label{eq:realspace_exchange}
\end{align}
where 
\begin{align}
    D_{k l}^{\sigma}(\mathbf{R})=\frac{1}{N_\mathbf{k}}\sum_\mathbf{k}\sum_{m} f_{m \sigma}(\mathbf{k}) c_{m \sigma}^{k}(\mathbf{k}) c_{m \sigma}^{l *}(\mathbf{k}) e^{i\mathbf{kR}}
    \label{eq:realspace_dm}
\end{align}
is the Fourier transform of the density matrix. $\mathbf{R}$ and $\mathbf{Q}$ denote lattice vectors; the sum over them is not restricted to the extent of the Born-von Karman cell, but solely by the overlap of the basis functions. The Greek symbols $\mu$ and $\nu$ are the indices of the auxiliary basis functions, as introduced next. The RI-LVL expansion is restricted in such a way that products of basis functions $\phi_{i}(\mathbf{r})$ at atom $I$ and $\phi_{j}(\mathbf{r})$ at atom $J$ are expanded in terms of auxiliary basis functions $P_\mu(\mathbf{r})$, which must be associated with the same pair of atoms $\mathcal{P}(IJ)$:
\begin{align}
    \phi_{i}(\mathbf{r}) \phi_{j}(\mathbf{r})=\sum_{\mu} C_{i j}^{\mu} P_{\mu}(\mathbf{r}).
\end{align}
Formally, this leads to demanding
\begin{align}
    C_{i j}^{\mu}=0, \quad \text{for } \mu \notin \mathcal{P}(I J),
\end{align}
a condition that can be fulfilled as the auxiliary basis set associated with $\mathcal{P}(IJ)$ approaches completeness.
Then, the RI expansion coefficients $C_{i k\left(\mathbf{R}\right)}^{\mu(\mathbf{Q})}$ can be derived as:
\begin{align}
    C_{i j}^{\mu}=\sum_{\nu \in \mathcal{P}(I J)}(i j \vert \nu) L_{\nu \mu}^{IJ},
    \label{eq:ri_coeffs}
\end{align}
with 
\begin{align}
    (i j \vert \nu)=\iint \phi_{i}(\mathbf{r}) \phi_{j}(\mathbf{r}) P_{v}\left(\mathbf{r}^{\prime}\right)v(\mathbf{r}-\mathbf{r}^{\prime}) d \mathbf{r} d \mathbf{r}^{\prime}
\end{align}
and the inverse Coulomb matrix $L_{v \mu}^{I J}=\left(V_{\mu \nu}^{I J}\right)^{-1}$ with 
\begin{align}
    V_{\mu \nu}(\mathbf{r}) =
    \begin{cases}
     \iint P_{\mu}(\mathbf{r}) P_{\nu}\left(\mathbf{r}^{\prime}\right) v(\mathbf{r}-\mathbf{r}^{\prime}) d \mathbf{r} d \mathbf{r}^{\prime}, \text{if }  \mu,\nu \in \mathcal{P}(I J)\\
     0, \text{otherwise}
    \end{cases}
    \label{eq:coulomb_matrix}
\end{align}
where $v(\mathbf{r}-\mathbf{r}^{\prime})$ is the Coulomb kernel, whose form depends on the chosen range-separation approach,~i.e.,~full-range for Hartree-Fock exchange, short- and long-range for range-separated hybrid exchange as defined in Eq.~\eqref{eq:coulomb_kernel}. 

For the actual implementation, the RI coefficients $C$ in Eq.~\eqref{eq:realspace_exchange} are grouped according to which atom pairs the auxiliary basis functions belong to. Following Ref.~\onlinecite{ihrig2015}, the exchange matrix for each pair of atoms $A_{1}^{1} A_{1}^{2}$ can be written as:
\begin{align}
X_{i \in A_{1}^{1} j \in A_{1}^{2}}^{\sigma}(\mathbf{R})=& \sum_{A_{2}^{1}\left(\mathbf{R}^{\prime}\right)} \sum_{k \in A_{2}^{1}\left(\mathbf{R}^{\prime}\right)} \sum_{v \in A_{1}^{2}} F_{i k}^{v(\mathbf{R})} E_{j k\left(\mathbf{R}-\mathbf{R}^{\prime}\right)}^{v \sigma} \nonumber \\
&+\sum_{A_{2}^{2}\left(\mathbf{R}^{\prime \prime}\right)} \sum_{l, v \in A_{2}^{2}\left(\mathbf{R}^{\prime \prime}\right)}\left(2 G_{i l}^{v \sigma}\left(\mathbf{R}+\mathbf{R}^{\prime \prime}\right)\right. \nonumber \\
&\left.+H_{i l}^{v \sigma}\left(\mathbf{R}+\mathbf{R}^{\prime \prime}\right)\right) C_{l j\left(-\mathbf{R}^{\prime \prime}\right)}^{v},
\label{eq:ri_lvl_exchange_matrix}
\end{align}
where
\begin{subequations}
\label{eq:temp}
\begin{align}
    E_{j k(\mathbf{R})}^{v \sigma}=& \sum_{I \mathbf{R}^{\prime \prime}} C_{j l\left(\mathbf{R}^{\prime \prime}\right)}^{\nu} D_{k l}^{\sigma}\left(\mathbf{R}+\mathbf{R}^{\prime \prime}\right), \label{eq:temp_E}\\
    F_{i k\left(\mathbf{R}^{\prime}\right)}^{\nu(\mathbf{R})}=& \sum_{\mu \in A(i)} C_{i k\left(\mathbf{R}^{\prime}\right)}^{\mu} V_{\mu \nu(\mathbf{R})}, \label{eq:temp_F}\\
    G_{i l}^{v \sigma}(\mathbf{R})=& \sum_{\mu \in A(i)}\left(E_{i l(-\mathbf{R})}^{v \sigma}\right)^{*} V_{\mu \nu(\mathbf{R})}, \label{eq:temp_G}\\
    H_{i l}^{v \sigma}(\mathbf{R})=& \sum_{k \mathbf{R}^{\prime}} F_{k i\left(-\mathbf{R}^{\prime}\right)}^{\nu\left(\mathbf{R}-\mathbf{R}^{\prime}\right)} D_{k l}^{\sigma}\left(\mathbf{R}-\mathbf{R}^{\prime}\right). \label{eq:temp_H}
\end{align}
\end{subequations}

\begin{figure}
    \centering
    \includegraphics[width=0.4\textwidth]{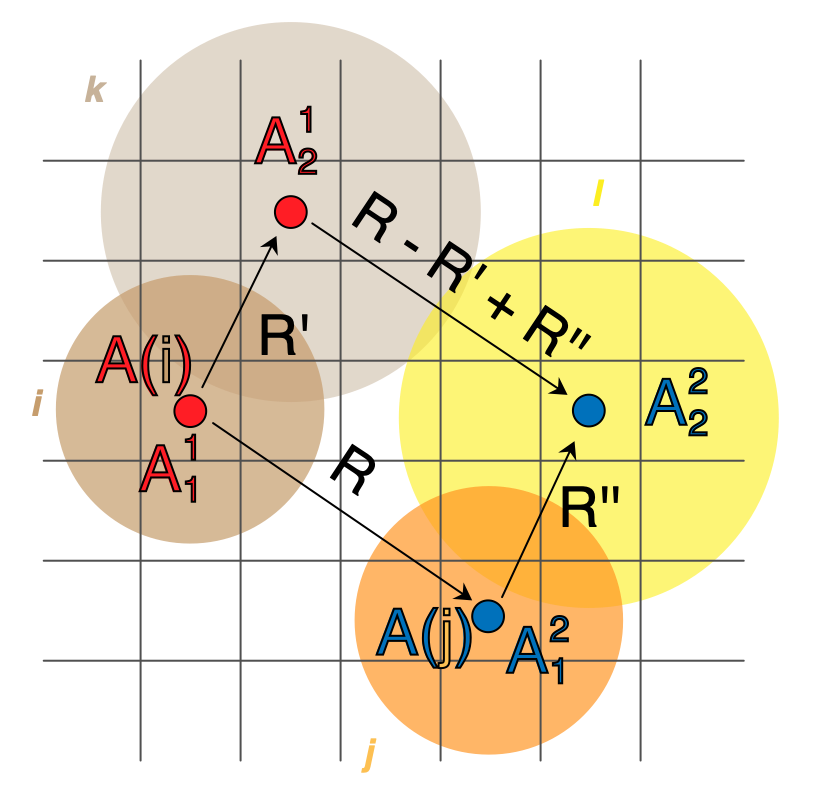}
    \caption{The labeling of the four atom centers and the lattice vectors connecting them used for grouping the RI-LVL four-center integrals in Eqs.~\eqref{eq:temp}. This figure is adapted from the original reference by S. Levchenko {\it et al.}~\cite{levchenko2015}.}
    \label{fig:four_center}
\end{figure}

As shown in Fig.~\ref{fig:four_center}, $A(i)$ denotes the atom on which basis function $i$ resides, $\mu \in A(i)$ signifies that the auxiliary function $P_{\mu}(\mathbf{r})$ is centered on atom $A(i)$. The $\ast$ symbol denotes complex conjugation. The choice of these intermediate matrices is motivated by the desire to minimize the number of matrix multiplications. The above equation makes use of the translational symmetry of the RI coefficients, namely $C_{i(\mathbf{0}) k\left(\mathbf{R}^{\prime}\right)}^{\mu\left(\mathbf{R}^{\prime}\right)}=C_{k(\mathbf{0}) i\left(-\mathbf{R}^{\prime}\right)}^{\mu(\mathbf{0})}$.

The above expressions can be extended to also allow for the computation of derivatives, namely the force and stress contributions stemming from Fock exchange. Details of these contributions can
be found in the original publication by Knuth {\it et al.}~\onlinecite{knuth2015all}.

\section{Description of the Algorithm and its Improvements}

\subsection{General concepts}
FHI-aims purely relies on the message passing interface (MPI) standard for parallelization. This choice has the advantage that any code implemented in this way will immediately work in parallel across multiple compute nodes, leaving the details of intra- vs. cross-node communication of data arrays up to the underlying MPI library. In recent years, however, the number of CPU cores per compute node increased faster 
than the total available memory per node. Therefore, arrays that are needed on all MPI tasks can significantly increase memory consumption on such architectures. To address this issue, a key change in our implementation was to move any large, precomputed coefficient arrays, e.g. the RI coefficients~$C_{i j}^{\mu}$ defined in Eq.~\eqref{eq:ri_coeffs} and the Coulomb matrix~$V_{\mu \nu}(\mathbf{r})$ defined in~Eq.~\eqref{eq:coulomb_matrix}, that are needed by all MPI tasks to shared memory arrays that are managed according to the MPI-3 standard (i.e., by the MPI library itself).
This choice improves scalability and reduces memory consumption significantly, since only one copy per node instead of one copy per core is stored in memory. For example, on a two-socket system with Intel Xeon IceLake-SP Platinum 8360Y CPUs (36 CPUs per socket, i.e. 72 CPUs per node), a reduction of the memory consumption for those arrays by roughly 2 orders of magnitude is achieved due this strategy alone.

Furthermore, one can exploit the fact that not all atom pairs have a significant overlap of basis functions, especially for large systems. Thus, a large number of computations and the associated memory cost can be avoided from the start. However, exploiting this sparsity requires a considerable bookkeeping effort to efficiently store, exchange, and use the sparsified data. In our optimized implementation of the EXX matrix computation, we store the global bookkeeping data for the various sparse arrays in MPI-3 shared memory arrays. By this means, all MPI tasks have access to the complete metadata (MPI task and offset) and can hence access the initialized arrays and the computed results via one-sided MPI calls.

One additional advantage of the described code infrastructure is that it facilitates efficient data reshuffling. We repeatedly exploit this property to optimize the data layout for the different stages of the computation. The initialization of the Coulomb and overlap matrices, the density computation, the actual EXX matrix computation, and the Pulay mixing and storage are all performed with a different data distribution to speed up computations and to reduce load imbalance at the various stages.

Moreover, this infrastructure allows to introduce an additional parallelization layer. In the 2015 implementation, the computation was only parallelized across atom pairs and the number of unit cells in the Born-von Karman cell, i.e., the set of real-space unit cells within which the Bloch phases of a finite, $\Gamma$-point centered $\mathbf{k}$-space grid are not yet periodically repeated. This restricted, coarse-grained parallelization inherently limits the scaling for large core counts because of the amount of data that needs to be exchanged. By decoupling the data layout from the actual work, the computations of the different $j$ columns of the exchange matrix can now be done independently, even if all of them require the data computed during initialization. Depending on the available memory, we evenly split the global MPI communicator into $n$ identical subcommunicators. We refer to these subcommunicators as \emph{instances} in the following. All of those instances are set up with everything that is necessary to compute any column of the exchange matrix Eq.~\eqref{eq:ri_lvl_exchange_matrix},~i.e.,~all of them have access to the precomputed data (e.g. RI coefficients and Coulomb matrix), the communicators for parallelization across atom pairs, and temporary arrays. This allows for the computation of different chunks of the exchange matrix $X_{ij}^{\sigma}(\mathbf{R})$ (Eq.~\eqref{eq:ri_lvl_exchange_matrix}) (called \emph{blocks} in the following) to be performed independently by different instances. The gathering of all blocks of the exchange matrix at the end is very fast compared to the computation. This additional parallelization layer drastically improves scalability: when enough memory is available, multiple instances can be spawned, and the strong scaling behaviour is significantly improved, as shown below.

In Fig.~\ref{fig:fockworkflow}, we show a sketch of the new workflow  for four compute nodes and a situation in which two instances are opened and three EXX matrix blocks are assigned to each of the two instances. Note that this distribution can change along the self-consistent field (SCF) convergence to adjust the load balance. The number of EXX matrix blocks is given by the number of basis functions ~$n_\text{basis}/\text{block size}$. We explain the individual steps in more detail in the following two subsections.

\subsection{Initialization}
\label{SSecInit}

The compute workflow and a schematic data layout for the initialization is sketched in the upper half of Fig.~\ref{fig:fockworkflow}. At the start of any calculation, the Coulomb matrix~$V_{\mu \nu}(\mathbf{r})$ as defined in~Eq.~\eqref{eq:coulomb_matrix} and the RI coefficients~$C_{i j}^{\mu}$ as defined in Eq.~\eqref{eq:ri_coeffs} are computed. Compared to the 2015 version, data re-use, memory access patterns, and vectorization have been improved. The parallelization for the initialization routines is updated so as to minimize load imbalance. As mentioned above, the arrays are then later redistributed and copied onto each instance to match the data layout of the computation of the EXX matrix in Eq.~\eqref{eq:realspace_exchange} during each SCF iteration, as indicated by the orange and green arrows in Fig.~\ref{fig:fockworkflow}. 
Furthermore, the usage of data compression was extended: In the 2015 implementation, only the Coulomb matrix was compressed by removing those columns and rows that exclusively feature elements with absolute values below a threshold of $10^{-10}$. The same compression method is now also used for the RI coefficients. The threshold was carefully tested so as not to alter the result to a numerically significant degree for both the Coulomb matrix and the RI coefficients.
The number of instances for the main computation is also determined during initialization. To this end, the available memory per node is measured and compared to the estimated memory consumption per instance to avoid out-of-memory situations. The memory consumption per instance is estimated based on the size of the Coulomb matrix and RI coefficients arrays, as well as heuristics for the largest temporary arrays during the main loop as defined by Eqs.~\eqref{eq:temp}. We use the following formula to estimate the ideal number of instances,~i.e.,~the number that gives the best performance and still fits into memory,
\begin{align}
    n_\text{instances} = \frac{n_\text{nodes}\times\text{min}(M_\text{free}-M_\text{buffer})}{M_\text{Coulomb}+M_\text{RI} + M_\text{main}},
    \label{eq:n_instances}
\end{align}
with the total number of nodes $n_\text{nodes}$, the currently available memory per node  $M_\text{free}$, the estimated memory usage needed for the actual evaluation per node $M_\text{main}$, the memory buffer per node $M_\text{buffer}$, the memory for the Coulomb matrix $M_\text{Coulomb}$, and the memory for the RI coefficients $M_\text{RI}$. However, only divisors of the total number of nodes are allowed (e.g. for 4 nodes only 1, 2, or 4 instances can be created), or more than one instance per node is also possible. With this mechanism, the code can use the entire available system memory over a wide range of nodes. In a strong-scaling scenario, the increasing total memory available will lead to an increasing number of instances, leading to a virtually perfect scaling of the Fock matrix computation, as long as there are enough blocks to distribute.
When the memory requirements of the expected computation are high compared to the available memory, only one instance spanning all MPI tasks will be created. By this means, data redundancy is avoided and the memory is used as efficiently as possible. Effectively, this recovers the original parallelization scheme used in the 2015 version, but the remaining computational benefits of the more fine-grained load balancing and of the MPI-3 shared memory arrays still lead to a significant performance improvement. Eventually, the Coulomb matrix and RI coefficients are copied to and redistributed on each instance to achieve optimal performance during the main loop of the calculation.
 
As discussed above, the computation of the real-space EXX exchange matrix, Eq.~\eqref{eq:realspace_exchange}, is performed in blocks over the last index $j$, which improves cache usage and reduces the number of MPI calls within an instance. We refer to this blocking of the Fock matrix as Fock matrix blocks or simply blocks in the following. The optimal block size is determined during initialization, but can also be set manually as an input parameter. Since many of the temporary arrays scale with the block size, this quantity has a considerable impact on the memory consumption per instance. During the initialization, the blocks of the exact exchange matrix are distributed evenly across all instances. During the main computation, the number of blocks per instance is increased/lowered after each SCF cycle according to the actual runtimes for the individual blocks to achieve optimal load balance. 

In the remainder of the paper, we will collectively refer to the determination of number of instances, the determination of the Fock matrix block size, and the redistribution of the Fock matrix blocks as auto-tuning mechanisms. 

\subsection{Evaluation of the EXX matrix in real space}
\begin{figure}
    \centering
    \includegraphics[width=0.45\textwidth]{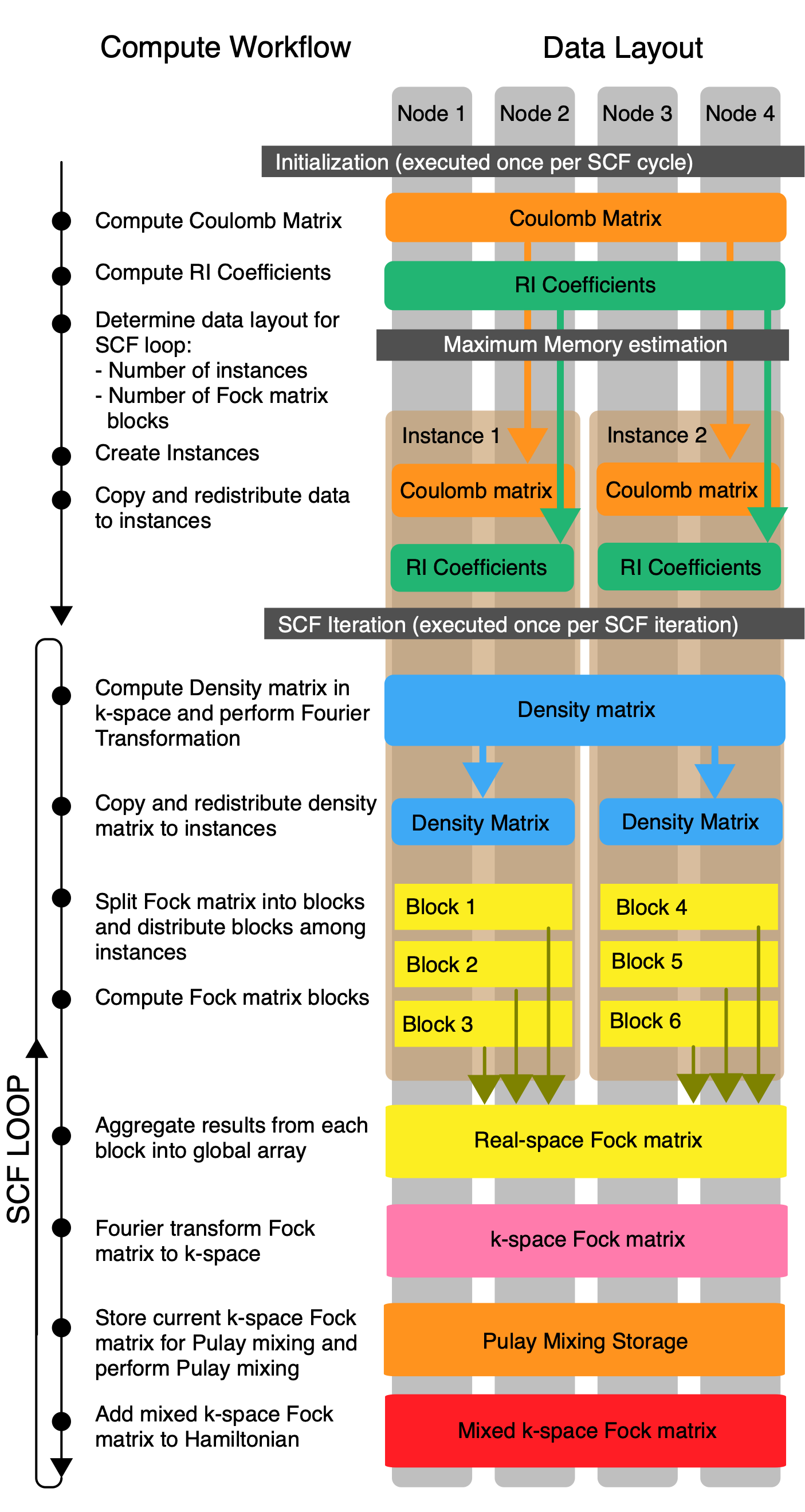}
    \caption{The computational workflow and data layout for the Fock matrix computation for the optimized algorithm. The data layout is shown here for a calculation that uses four nodes, two instances of the computational infrastructure (see text) and three blocks of the EXX matrix are treated by each instance.}
    \label{fig:fockworkflow}
\end{figure}

The EXX matrix is evaluated once per SCF iteration. The process is outlined in the lower half of Fig.~\ref{fig:fockworkflow}, where the left column describes the individual steps that are executed and the right column indicates the data layout used for the largest arrays. As a first step, the (un-mixed) density matrix is constructed from the eigenvectors of the previous solution of the KS eigenvalue problem, and is Fourier transformed into real space to obtain $D_{k l}^{\sigma}(\mathbf{R})$ according to Eq.~\eqref{eq:realspace_dm}. For the first SCF iteration, the solution of the KS eigenvalue problem for the semilocal PBE functional with an initial density of superimposed spherical free atoms or ions is used, although more sophisticated choices could be pursued in future work. For all subsequent SCF iterations, the density matrix from the solution of the KS eigenvalue problem using the actual hybrid functional is used. The data layout and communication patterns for the Fourier transforms in Eqs.~\eqref{eq:fock_operator_k} and ~\eqref{eq:realspace_dm} have been optimized for different stages in the computation workflow: E.g., the computation and Fourier transformation of the distributed density matrix is first computed efficiently across all MPI tasks and, for the subsequent computation steps, the real-space density matrix~$D_{k l}^{\sigma}(\mathbf{R})$ is redistributed and stored in a different data layout to be optimal for the matrix multiplications in Eqs. \eqref{eq:temp_E} and \eqref{eq:temp_H}, as indicated by the blue arrows in Fig.~\ref{fig:fockworkflow}. 

In the following we briefly outline the computation of a row of the EXX matrix Eq.~\eqref{eq:realspace_exchange} in pseudocode. The notation for the temporary matrices is the same as introduced in Eqs.~\eqref{eq:temp}. \verb|C| refers to the RI coefficients Eq.~\eqref{eq:ri_coeffs}, while \verb|C'| refers to reordered RI coefficients to compute Eq.~\eqref{eq:temp_G} efficiently. \verb|D| refers to the density matrix in real space Eq.~\eqref{eq:realspace_dm}. The variable names used on the left hand side of the pseudo-instructions reflect their naming in the actual code.
The workflow and the data layout that is used during the simulation is outlined in Fig.~\ref{fig:fockworkflow}.
\begin{verbatim}
    compute RI coefficients * density matrix:
        tmp             := E  = C*D
    sum up first part of EXX matrix:
        temp_prod       := F  = C*V
        fock_matrix_mem := X += temp_prod*tmp
    compute remaining temporary arrays: 
        tmp2            := H  = F*D
        tmpx2           := E' = C'*D
        temp_prod       := G  = tmpx2*V
        tmp2            += 2*temp_prod
    compute second part of EXX matrix:
        fock_matrix_row += C*tmp2
        fock_matrix_row += fock_matrix_mem
\end{verbatim}

We refer to a collection of several EXX matrix rows as a \emph{block} in the following, as also shown by the yellow boxes in Fig.~\ref{fig:fockworkflow}. During the first SCF step, all blocks are distributed evenly among the instances, but they might be later redistributed to optimize the load balance. In order to use the available memory efficiently, several temporary arrays are re-used (and over-written) during the computation of a block. Whether or not temporary arrays or a product of temporary arrays are computed is decided based on estimates of the maximum norms of some of the arrays involved, e.g., the density matrix, or the RI coefficients. These screening mechanisms have not changed and are described in detail in the 2015 paper~\cite{levchenko2015}.

For systems in which not all atom pairs overlap, dynamic load imbalance occurs in the main loop running over the global basis function index \texttt{n\_basis}  due to the parallelization of the coulomb/overlap matrices over \texttt{n\_atoms}. This means that the relative computational load for different MPI tasks varies between SCF iterations because the work required for a particular basis function with index~\texttt{i\_basis} depends on the overlap between the atom associated with \texttt{i\_basis} and the atom associated with the data stored locally on the task and on the synchronization points (global collectives) in each iteration. We implemented blocking to reduce the number of synchronization points. Note that the indices \texttt{i\_basis} accessed in each block are not consecutive,~i.e.,~not associated with the same atom, but aim to achieve some balancing between all MPI tasks within each block.

After the EXX matrix is fully constructed in real space, it is backtransformed into reciprocal space (pink box in Fig.~\ref{fig:fockworkflow}). The Pulay mixing algorithm~\cite{PULAY1980} with mixing factors based on the density is used as a default to achieve efficient SCF convergence. As described initially, we use the unmixed density matrix to construct the Fock matrix. Subsequently, we apply the Pulay mixing factors of the density to the EXX exchange matrix as well. This choice significantly increases the memory consumption for large-scale systems, since, by default, the previous eight EXX matrices are stored and all of them are of the size of the Hamiltonian: $n_\text{basis}\times n_\text{basis}$ for each $\mathbf{k}$ point (orange box in Fig.~\ref{fig:fockworkflow}). We only store EXX matrices for the set of non-equivalent $\mathbf{k}$-points per time-reversal symmetry, unless otherwise requested by the user. This saves a factor of about two for dense k-grids. Also, the data layout has been optimized for storing the current EXX matrix while executing the Fourier back transform. Eventually, the corresponding fraction of the Pulay mixed k-space EXX matrix is added to the Hamiltonian.

\subsection{EXX force and stress contributions}
\label{SSecForceStress}

FHI-aims also provides analytical forces and stress evaluation for hybrid DFAs~\cite{knuth2015all}, which enable, e.g., (periodic) structure optimization or molecular dynamics. For the theoretical background, we refer to the original paper~\cite{knuth2015all}. The Fock exchange force contribution is only computed once per SCF cycle in an additional SCF step when the electronic convergence criteria have been reached. Each evaluation of the three force components and each evaluation of the six (or nine) stress tensor components require a computation of the same size and complexity as the original Fock matrix. In a na\"ive implementation of these Fock derivatives, the total memory consumption and runtime would increase by a factor of four, if only forces are computed, or by a factor of ten if also the stress is computed. Especially for large systems, this factor would lead to a huge and undesirable memory consumption. Therefore, we implemented a splitting of the computation of the stress components into several parts. The code determines the mode based on the available remaining memory. If the problem size is small, all force and stress components are computed in parallel. If the problem size is large, the force and stress computation is split into three parts: first, the exchange matrix plus the three force components are computed, second, the first three stress components, and, in a third part, the remaining three stress components are computed. This choice keeps the memory consumption manageable; however, it increases the computation time by a factor of two. 

\subsection{Improvements of other relevant code parts}

In addition to the EXX matrix evaluation, the evaluation of the electrostatic (Hartree) potential and of the Pulay force terms was also optimized to reduce the computational time spent for large periodic systems. This improvement benefits both semilocal and hybrid DFT computations. In FHI-aims, the Hartree potential in each SCF step is evaluated as a difference to the sum of free atom potentials, $\delta v_\mathrm{es}(\mathbf{r})$~\cite{blum2009}. The difference is computed on each point of the integration grid by summing up atom centered multipole components $\delta \tilde{v}_{\mathrm{at},lm} (r)$ for each atom, see Ref.~\onlinecite{blum2009}. In our implementation, we restructured the computations such that for each multipole component its contribution to the Hartree potential is evaluated for a batch of points. This avoids branching in the innermost loop and reduces the number of subroutine calls by two orders of magnitude, improves memory accesses and cache reuse and allows for compiler vectorization.

For periodic systems, the Hartree potential is evaluated using the Ewald method that decomposes the potential into short- and long-range components. For the long-range components, we introduced a blocking method that evaluates the potential for a batch of points. Here it was possible to rewrite the computations using highly tuned dgemm/zgemm routines of the BLAS (Basic Linear Algebra Subprograms) standard library instead of a loop based implementation, which greatly improved the computational efficiency.

Similar to the evaluation of the Hartree Potential, we simplified the branching inside of the main loops of the Pulay force evaluation and restructured the computation to aggregate/avoid unnecessary computations. In addition, we improved the repeated initialization of some large arrays in one of the main loops by exploiting their sparsity, which also helped to reduce computation time significantly.

\section{Benchmark results}
\label{SecBenchmark}

\subsection{Benchmark platform (hardware and software specification)}

The benchmark calculations reported here were performed on the Lenovo HPC system \textit{Raven} \cite{MPCDF_Raven_doc} at the Max Planck Computing and Data Facility. The compute nodes provide two Intel Xeon Platinum 8360Y (IceLake-SP) processors with 72 cores per node, operated at the nominal frequency of 2.4 GHz (turbo mode disabled). Depending on the memory (RAM) requirements, we either use nodes equipped with 256 GB RAM or 512 GB RAM. Both types of nodes share the same memory-performance characteristics (ca. 310 GB/s sustained bandwidth on the stream triad microbenchmark \cite{McCalpin1995}).  
All nodes are connected through a 100 Gb/s Nvidia Mellanox HDR100 InfiniBand network with a non-blocking fat-tree topology. We use the Intel ifort compiler 2021.6.0, the Intel MPI library 2021.6, and the Intel MKL library 2022.1.

\subsection{Strong and weak scaling behavior}
In the following we compare the parallel scaling behavior of the new implementation with the original version from 2015. We show the $O(N)$ scaling with respect to increasing the number of atoms in the system, while keeping the number of basis functions per atom constant.  

We compare the performance of the improved code to the original implementation from 2015~\cite{levchenko2015} for GaAs supercells of different sizes. We use the $4\times4\times4$ primitive unit cell as 128-atom supercell and subsequently double the cell in each direction, so we get supercell sizes of 256, 512, and 1024 atoms. The k-grid for the 1024-atoms supercell is chosen as $1\times1\times1$, and the k-grid density is kept consistent for the remaining supercells. We use the \emph{intermediate~2020} species defaults of FHI-aims (which are the earlier "tight" settings in the 2010 notation as used in the 2015 implementation). The calculations are all-electron calculations with no shape approximations to the underlying electron-nuclear potential and they are carried out without making use of any space group symmetries or other system-specific simplifications.

\begin{figure}
    \centering
    \includegraphics[width=0.49\textwidth]{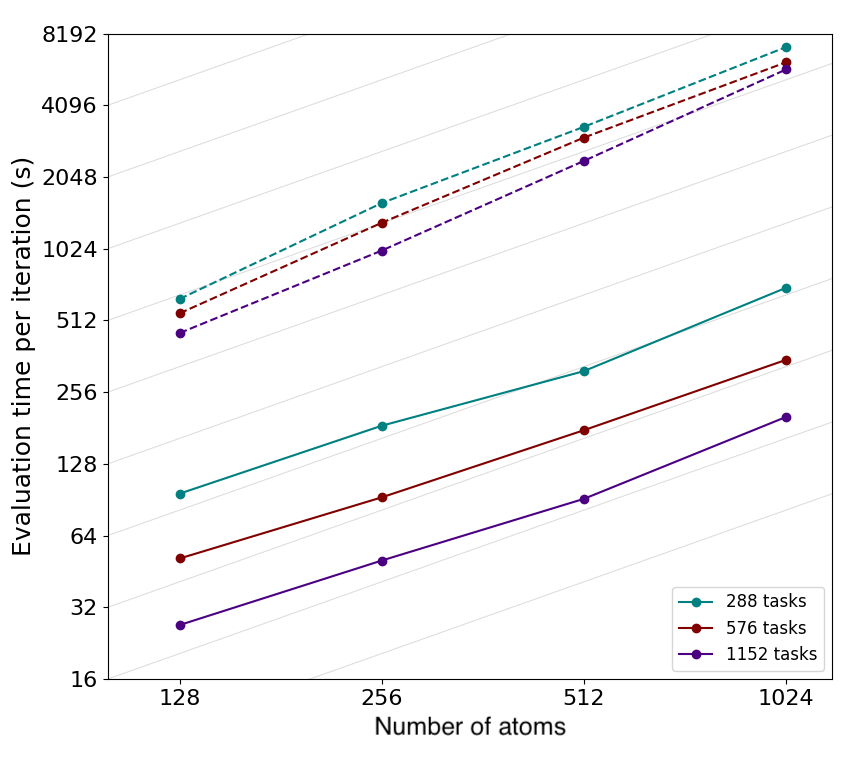}
    \caption{The O(N) scaling of the improved (solid lines) and the 2015 (dashed lines) implementation of the HSE06 exchange evaluation timings per SCF iteration for GaAs supercells with 256, 512, and 1024 atoms using the {\it intermediate} FHI-aims species defaults, i.e., 34 NAO basis functions per atom. The grey lines indicate linear scaling. The calculations were run on nodes with 256 GB RAM.}
    \label{fig:order_N_scaling}
\end{figure}

\paragraph{Strong scaling}
The drastic improvements in the code performance are showcased in Fig.~\ref{fig:order_N_scaling}, which presents actual runtimes for one evaluation of the EXX (in the original and the improved implementation) for each GaAs supercell as a function of the number of atoms included in the supercell. Regardless of the number of nodes used, a huge reduction in the runtime – roughly at least an order of magnitude – is observed for all system sizes considered. We observe a dependence of the runtime that approximately scales as $aN^b$ with system size~$N$. This scaling reveals that the increase in performance is rooted in both a massive reduction of the prefactor $a$ and in improvements of the scaling exponent $b$. For small system sizes, the observed reduction in the prefactor by about a factor of 6 (4 nodes) and 9 (16 nodes) is largely responsible for the obtained savings. For larger system sizes, the reduced scaling exponent becomes more important: Compared to the original implementation, which already features a favorable, almost linear scaling with $b\approx1.1-1.26$, the improvements resulted in a further decrease of the scaling coefficient, leading to an almost perfect linear scaling ($b \approx 1.01-1.05$) of the computational cost with $N$. This is remarkable, since such a consistent linear scaling is extremely hard to achieve in actual implementations, and this is already observed for the relatively moderate system sizes considered in this plot ($N \leq 1024$ atoms).

\begin{figure}
    \centering
    \includegraphics[width=0.49\textwidth]{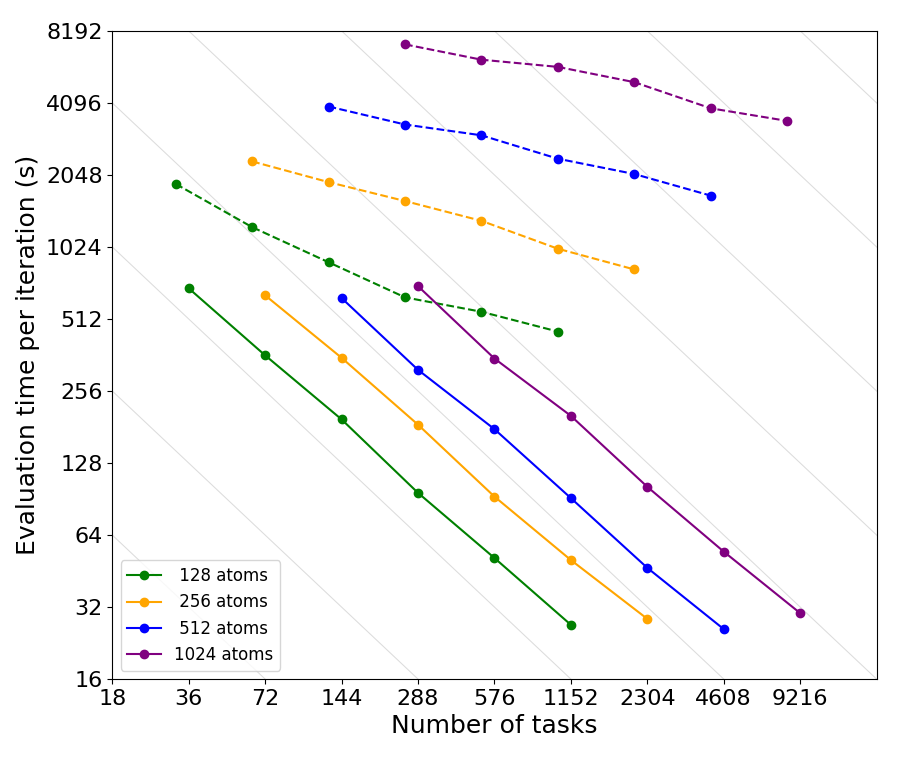}
    \caption{Strong scaling of the improved (solid lines) and the 2015 (dashed lines) implementation of the HSE06 exchange evaluation timings per SCF iteration for GaAs supercells with 256, 512, and 1024 atoms using the {\it intermediate} FHI-aims species defaults, i.e., 34 NAO basis functions per atom. The grey lines indicate ideal strong scaling. The calculations were run on nodes with 256 GB RAM.}
    \label{fig:strong_scaling}
\end{figure}

The observed increase in the code performance also translates into a better strong scaling behavior, as showcased in Fig.~\ref{fig:strong_scaling}. This plot shows actual runtimes for one evaluation of the exact-exchange operator as a function of the number of nodes viz. cores. More specifically, GaAs supercells with 256, 512, and 1024 atoms were investigated using both the original implementation and the one including our improvements. As mentioned when discussing Fig.~\ref{fig:order_N_scaling}, we already observe large improvements in the runtime of roughly one order of magnitude for small node counts. The improvements are even more pronounced for larger-scale calculations featuring thousands of cores and can reach reductions of the runtime by two orders of magnitude and even more. Compared to the original implementation, in which the runtime~$t(n)\approx 1/n^c$ scales with respect to the number of cores, $n$, with an exponent $c$ between 0.2 and 0.25, the improved implementation scales with an exponent~$c$ between 0.9 and 0.95, very close to an ideal speedup. Clearly, a more favorable scaling is observed for larger workloads viz. system sizes. As shown in the next section, the developed routines show a similarly good scaling on a much larger number of nodes and cores if the respective problem size is increased accordingly, 

\begin{figure}
    \centering
    \includegraphics[width=0.49\textwidth]{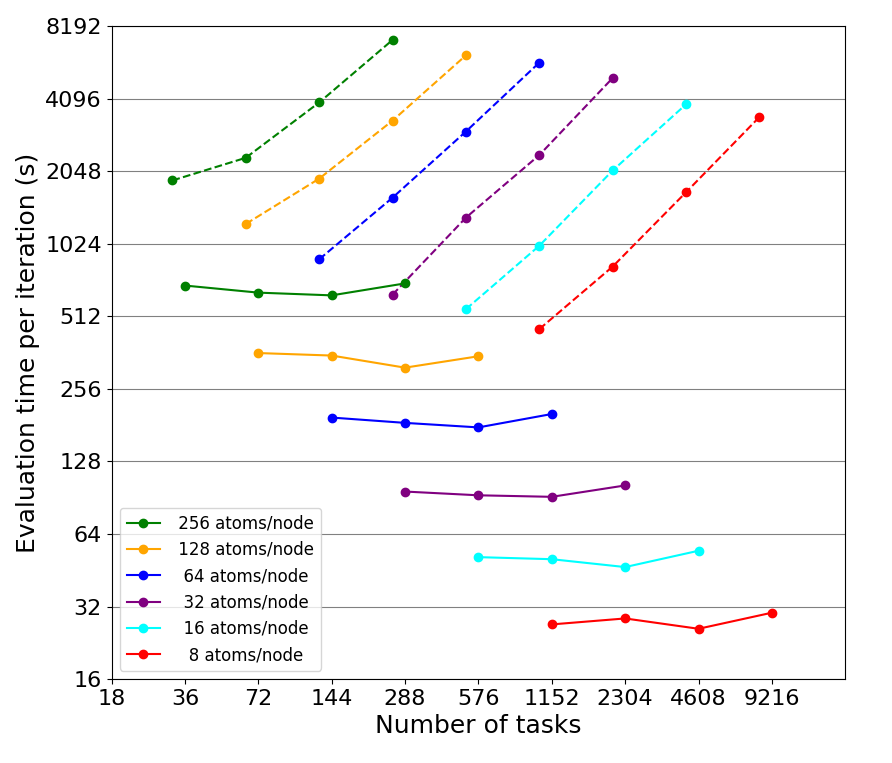}
    \caption{Weak scaling for the improved (solid lines) and 2015 (dashed lines) implementation of the HSE06 Fock exchange evaluation timings per SCF iteration for GaAs supercells with 256, 512, and 1024 atoms using the {\it intermediate} FHI-aims species defaults, i.e., 34 NAO basis functions per atom. The grey lines indicate ideal weak scaling. The calculations were run on nodes with 256 GB RAM.}
    \label{fig:weak_scaling}
\end{figure}

\paragraph{Weak scaling}
To showcase weak scaling with a constant workload per node, cf.~Fig.~\ref{fig:weak_scaling}, we use the same GaAs supercell data as in the previous section.  For the 2015 implementation, we observe that the average run time for a constant workload increases almost linearly with higher node counts, indicating a sub-optimal parallelization scheme. For the new implementation, this problem has been solved due to the introduction of the additional parallelization layers. As shown in Fig.~\ref{fig:weak_scaling}, the run time is close to constant for a given workload. 

\subsection{Benchmark calculations for selected systems}
Using the algorithm described above, FHI-aims can perform hybrid density functional calculations for both periodic and non-periodic systems. To test the limits of the implementation and document the current runtimes as a reference for future developments, we have selected different systems and geometries. In total, we addressed 18 systems (ten bulk materials, two surfaces, two nanosystems, two clusters, two molecules). The number of atoms in the systems range from two to more than 30,000 atoms. All details and references to the original publications discussing these systems as well as full access to the simulation results are given in the SI. In this work, we only focus on systems that exhibit a gap between the highest occupied and lowest unoccupied states. Testing the limits of the implementation with respect to the number of atoms has helped to identify and then resolve issues with order-$N_\text{atoms}^2$ arrays across different code parts in FHI-aims, which at some point would otherwise dominate the memory consumption. In addition to Tables~\ref{tab:bulk} and \ref{tab:nano_nonperiodic} and the SI, several key results are included in Figure~\ref{fig:large-scale-overview}.

Our test set covers both high- and low density compounds to explore the limits of the parallelization. In this context, it is important to keep in mind that the local atom density, i.e., the number of atoms within the overlap radius of the basis functions, defines the workload. Formally, the workload scales with $O(n_\text{basis}^4)$, where $n_\text{basis}$ is the number of basis functions with non-vanishing overlap.
 For instance, the simulated carbon allotropes are very dense, so the workload is comparatively high, although each atom has only a few basis functions. In contrast, the ice XI or paracetamol crystals have a low atom density and, in addition, also only few basis functions, so the workload is low. In addition, we also included systems that contain both heavy and light elements,~e.g.,~the inorganic/organic hybrid lead iodide perovskite methylammonium lead iodide (MAPI) or the layered hybrid perovskite phenethylammonium lead iodide (PEPI). These types of systems represent an important computational challenge, i.e., the workload per atom is very inhomogeneous since atoms from very light (H) to very heavy (Pb) elements are included. The refined parallelization scheme over basis functions as well as the establishment of runtime auto-tuning mechanisms that adapt the workload while running the calculations are critical for efficiently executing these calculations. Finally, we also included magnetic systems, i.e., the Fayalite and Hematite unit cells. Using collinear spin doubles the memory and runtime since the spin index in Eq.~\eqref{eq:realspace_exchange} runs up to two. 

FHI-aims provides pre-defined defaults for individual atoms that include the integration grids, basis functions and their spatial extent, as well as the expansion order of the mean-field electrostatic potential. In this work, we mostly use ``light'', ``intermediate'', and ``tight'' species defaults. We here use the defaults denoted as "2020 defaults" in FHI-aims. As the naming indicates, they represent defaults with increasing accuracy, but also higher computational costs. We note that the ``intermediate'' species defaults are the recommended production settings for hybrid DFT calculations in FHI-aims, such as, high quality geometry relaxations, sophisticated band structures and energy differences. In turn, the calculations in Tables.~\ref{tab:bulk}, ~\ref{tab:nano_nonperiodic}, ~\ref{tab:force_stress}, and ~\ref{tab:PBE_PBE0_HSE06} with intermediate species defaults can be considered as converged and indicative of realistic simulation settings for DFT simulations of the  considered materials. For  periodic systems, we also used k-grid densites that can be considered fully converged for systems that provide a gap. We use a k-grid density of $n_i \cdot a_i > 50~\text{\AA}$, where $a_i$ is the lattice vector length and  $n_i$ is the number of k-points along the corresponding k-space direction~$i$. The k-point grids are deliberately chosen to be very well converged, illustrating that the current implementation can be used without \emph{a priori} precision tradeoffs in this respect. For example, even for the 2,000 atom diamond carbon supercell a $2\times 2\times 2$ k-space grid is used in our benchmark.
We emphasize that for most systems even the less computationally demanding ``light'' settings would lead to qualitatively correct and quantitatively acceptable results, e.g. for geometry optimization or band gap computations.
Likewise, somewhat reduced k-grids would lead to notably lower computational cost and results that can be deemed acceptable. We further note that all calculations shown here are using  runtime parameters, e.g., the number of instances and the Fock matrix block size, that are automatically chosen by the auto-tuning mechanism. As discussed in Sec.~\ref{SSecInit}, this technique aims at yielding very good computational performance by default and to concurrently avoid out-of-memory scenarios, since conservative memory estimates are used. Accordingly,  a manual fine tuning of the parameters for the number of instances and the Fock matrix block size can further speed up calculations by roughly another 25\% for most systems.

\paragraph{Bulk systems.} The bulk system class is computationally the most challenging one. Due to the 3D periodicity, the neighboring unit cells in the Born-von Karman cell lead to contributions for the Fock-type exchange evaluation in all directions. Boron nitride and the carbon diamond supercell are the densest materials among the bulk test suite with about 0.17 and 0.18 atoms/\AA$^{3}$, respectively. As discussed earlier, despite a low number of basis functions per atom, this leads to a potentially high workload per basis function since the high atom density increases the number of overlapping basis functions from the neighboring atoms. As a result, the computational effort scales as $O(N_b^4)$ with $N_b$ the number of nonzero basis functions in a given volume element, since no sparsity can be exploited for basis functions that are placed on top of one another. In Table~\ref{tab:bulk}, we compare ``light'' and ``intermediate'' species defaults for most of the systems in the test suite, where intermediate settings are designed to provide sufficient numerical precision for any production DFT simulations. For boron nitride, we also include the ``tight'' species defaults, illustrating the scaling with the number of basis functions. For DFT production calculations, the intermediate species defaults are recommended, as most ground state observables (e.g. energy, forces, stress), but also the KS eigenvalues for band structures are well converged with these settings. The Hematite and Fayalite systems require collinear spin polarization to be included in the calculations in order to obtain the correct electronic structure for their ground states. Both systems have an antiferromagnetic spin ordering. The Hydrogen interstitial introduces a shallow defect level that is partially occupied. Among the largest structural models, we include supercells of cubic methylammonium lead iodide perovskite (MAPI) with 768 atoms, a defect complex (a lead vacancy with two Bi substitutions at the Pb site) in the phenylethylammonium lead iodide perovskite with 3,383 atoms~\cite{Lu2023}, a $4\times4\times4$ paracetamol supercell containing over 10,000 atoms. The largest system is a supercell of ice XI with over 30,000 atoms (Figure~\ref{fig:large-scale-overview}), treated with light species defaults, for which we used nodes with 512 GB memory. For some systems, we indicate the strong scaling behavior by doubling the number of nodes. We observe that the strong scaling for the EXX-contribution evaluation remains intact even for systems beyond 10,000 atoms. For systems for which the number of instances is greater than the number of nodes, more than one instance per node is used.

\paragraph{Surfaces.} The simulation of surface slabs is a special case of the 3D periodic systems, in which a vacuum region is inserted in one of the three spatial directions. We selected a hydrogen-passivated $6\sqrt{3}\times6\sqrt{3}$ silicon carbide slab with nine atomic layers and a graphene sheet at the top (1,648 atoms). For this example, the Fermi level is pinned close to the Dirac cone of the graphene sheet, but due to interaction with the slab the Dirac cone states are partially occupied. Additionally, we choose a (100)-oriented TiO$_2$ slab (3,456 atoms) that possesses no metallic surface states. Both systems are large and are demonstrated in simulations using significant resources (32 and 48 nodes, respectively, using intermediate settings). Nevertheless, resources of this kind are available in many high-performance computing centers across the globe and the system sizes shown enable rather realistic simulations of nanoscale processes across chemistry and materials science, without undue interactions across periodic supercell images.

\paragraph{Nanosystems.} Two different Carbon allotropes are chosen: A Carbon nanowire and a Carbon nanotube. Both systems are one-dimensional in the sense that vacuum has been added in two spatial directions and only one direction is actually periodic. The less bulk-like the simulated system, the lower the actual workload. This can be directly observed for both nanosystems having the same number of atoms, but a very different EXX evaluation time. Comparing for the same number of cores and same species defaults, the nanotube can be evaluated more than five times faster than the nanowire, since the nanotube can be considered a rolled 2D graphene sheet with no bulk-like volume. In contrast, the nanowire has a bulk-like core. Nonetheless, the 2,000-atom Carbon diamond bulk needs a five times longer evaluation time on the same number of nodes. 
For either nanosystem, quite modest resources are required in view of the system size,~i.e.,~eight and sixteen nodes, respectively.

\paragraph{Clusters and Molecules.} These systems do not have periodic boundary conditions, but, as initially mentioned, can still be simulated with the same code as described above. We have selected a part of a DNA molecule solvated in saline water containing overall 15,613 atoms, a water cluster shaped as a sphere (``water drop'') containing 1,800 atoms, a silicon wire containing 706 atoms and a charged Ac-Lys-Ala19-H+ molecule containing 220 atoms. The DNA system is a remarkably large systems for hybrid DFA based simulations at the numerical precision level of intermediate settings. While 128 nodes for the 15,613-atom solvated DNA molecule are large resources, being able to capture processes in such a system with the accuracy of a hybrid DFA at all is, again, a considerable success.

In general, we observe that the per-SCF-step evaluation of the EXX matrix shows close-to-ideal strong scaling across the systems. However, the EXX initialization timings do not always scale with the number of nodes. In practice, this has litte influence on overall runtimes, since the initialization is only evaluated once per SCF cycle and usually does not exceed the cost of an extra SCF step. 

\begin{table}[ht]
\centering
\begin{tabular}{|l|r|r|r|r|r|r|r|r|}
    \hline
    \multicolumn{9}{|c|}{\textbf{Bulk systems}} \\
    \hline
    Species & Nodes & Inst. & Init (s) & Fock (s) & KS (s) & \#Basis & \#States & M (GB) \\ \hline
     \multicolumn{9}{|c|}{Boron Nitride: BN (2 atoms), 19$\times$19$\times$19 k-grid} \\ \hline \hline
     light & 1 & 8 &   71.384 &    2.809 &    0.054 & 28 & 12 & 86.31 \\ \hline
     interm. & 1 & 8 &  115.769 &   20.084 &    0.128 & 60 & 12 & 131.34 \\ \hline
     tight & 1 & 3 &  147.644 &   84.480 &    0.389 & 78 & 12 & 158.19 \\ \hline
     \multicolumn{9}{|c|}{Carbon Diamond: C$_{2000}$ (2,000 atoms), 2$\times$2$\times$2 k-grid, Ref.~\onlinecite{kecceli2016}} \\ \hline \hline
     \multirow{2}{*}{light} & 16 & 1 & 45.677 &  130.690 & 60.236 & \multirow{2}{*}{28,000} & \multirow{2}{*}{12,000} & 142.85 \\
     & 32 & 4 & 42.704 &  53.065 & 34.734 & & & 141.69  \\ \hline
     interm. & 64 & 1 & 139.876 &  459.099 & 57.022 & 60,000 & 12,000 & 207.80\\ \hline
     \multicolumn{9}{|c|}{Cubic MAPI: C$_{64}$N$_{64}$H$_{384}$Pb$_{64}$I$_{192}$ (768 atoms), 2$\times$2$\times$2 k-grid, Ref.~\onlinecite{mohr2015}} \\ \hline \hline
     light & 8 & 2 &   68.575 &   69.039 &   27.557 & 15,744 & 10,912 & 97.92 \\ \hline
     \multirow{2}{*}{interm.} & 8 & 1 &  123.644 &  179.267 &   32.493 & \multirow{2}{*}{20,672} & \multirow{2}{*}{10,912} & 127.11 \\
     & 16 & 2 &   83.606 &   92.080 &   18.611 &  &  & 98.87 \\ \hline
     \multicolumn{9}{|c|}{Hematite: Fe$_4$O$_{6}$ (10 atoms), 10$\times$10$\times$10 k-grid, collinear spin, Ref.~\onlinecite{Villars2023:sm_isp_sd_0314193}} \\ \hline \hline
     light & 1 & 8 &   17.090 &   26.263 &    0.263 & 208 & 116 & 93.99 \\ \hline
     interm. & 1 & 3 &   48.562 &   83.854 &    0.410 & 340 & 116 & 93.99 \\ \hline
     \multicolumn{9}{|c|}{Fayalite: Fe$_8$Si$_4$O$_{16}$ (28 atoms), 4$\times$7$\times$8 k-grid, collinear spin, Ref.~\onlinecite{Villars2023:sm_isp_sd_0375064}} \\ \hline \hline
     light & 2 & 8 &   11.738 &   18.915 &    0.923 & 572 & 300 & 85.60 \\ \hline
     interm. & 2 & 2 &   36.663 &   92.804 &    1.544 & 936 & 300 & 93.18 \\ \hline
     \multicolumn{9}{|c|}{Hydrogen interstitial in Silicon: Si$_{64}$H$_1$ (65 atoms), 4$\times$4$\times$4 k-grid, Ref.~\onlinecite{lin2014siesta}} \\ \hline \hline
     light & 1 & 2 &   28.297 &   62.990 &    2.561 & 1,605 & 643 & 117.99 \\ \hline
     \multirow{2}{*}{interm.} & 1 & 1 &   80.690 &  225.392 &    2.588 & \multirow{2}{*}{2,187} & \multirow{2}{*}{643} & 179.04 \\
     & 2 & 2 &   44.763 &  112.293 &    1.337 &  & & 127.11 \\ \hline
     \multicolumn{9}{|c|}{Water (Liquid): H$_{128}$O$_{64}$ (192 atoms), 4$\times$4$\times$4 k-grid, Ref.~\onlinecite{lin2014siesta}} \\ \hline \hline
     light & 1 & 6 &   16.157 &    8.752 &    2.475 & 1,536 & 704 & 115.58 \\ \hline
     interm. & 1 & 1 &  102.727 &  101.951 &    4.975 & 3,328 & 704 & 193.93 \\ \hline
     \multicolumn{9}{|c|}{PEPI with Defect: C$_{1,152}$H$_{1,728}$Bi$_2$I$_{288}$N$_{144}$Pb$_{69}$ (3,383 atoms), 1$\times$1$\times$1 k-grid, Ref.~\onlinecite{Lu2023}} \\ \hline \hline
     \multirow{2}{*}{interm.}& 16 & 1 &  174.454 &  299.968 &   69.529 & \multirow{2}{*}{75,100} & \multirow{2}{*}{24,076} & 209.03 \\
      & 32 & 1 &  140.919 &  227.803 &   56.663 &  &  & 131.63 \\
     \hline
     \multicolumn{9}{|c|}{4$\times$4$\times$4 Paracetamol: C$_{4,096}$H$_{4,608}$N$_{512}$O$_{1,024}$ (10,240 atoms), 1$\times$1$\times$1 k-grid, Ref.~\onlinecite{rossi2016anharmonic}} \\ \hline \hline
     \multirow{2}{*}{light} & 16 & 1 &  216.257 &  323.327 &  380.769 & \multirow{2}{*}{101,888} & \multirow{2}{*}{44,288} & 214.16 \\
      & 32 & 4 &  207.386 &  106.763 &  218.647 & & & 205.84 \\ \hline
     interm. & 84 & 1 &  409.863 &  803.045 &  278.179 & 219,648 & 44,288 & 215.74\\ \hline
     \multicolumn{9}{|c|}{$^\ast$Supercell of Ice XI: H$_{20,384}$O$_{10,192}$ (30,576 atoms), 1$\times$1$\times$1 k-grid, Ref.~\onlinecite{Leadbetter1985}} \\ \hline \hline
     light & 60 & 5 & 1,418.280 &  710.858 & 1,670.647 & 244,608 & 112,112 & 443.27 \\ \hline
     \hline
     \multicolumn{9}{|c|}{\textbf{Surfaces}} \\ \hline
     \multicolumn{9}{|c|}{Graphene on SiC slab: C$_{1,108}$Si$_{432}$H$_{108}$ (1,648 atoms), 2$\times$2$\times$1 k-grid, Ref.~\onlinecite{nemec2013thermodynamic}} \\ \hline \hline
     light & 8 & 1 &   76.334 &  181.153 &   59.822 & 26,852 & 11,184 & 137.91 \\ \hline
     interm. & 32 & 1 &  107.521 &  516.838 &   32.500 & 49,116 & 11,184 & 141.12 \\
     \hline
     \multicolumn{9}{|c|}{TiO$_2$ slab: Ti$_{1,152}$O$_{2,304}$ (3,456 atoms), 1$\times$1$\times$1 k-grid} \\ \hline \hline
     light & 32 & 2 &   86.544 &  149.935 &   84.533 & 67,968 & 35,136 & 107.82 \\ \hline
     interm. & 48 & 1 &  176.155 &  615.024 &  113.114 & 115,200 & 35,136 & 175.15 \\
     \hline

\end{tabular}
\caption{All calculations are carried out with the HSE06($\alpha=$0,$\beta$=0.25,$\omega$=0.11 Bohr$^{-1}$) functional. Species: Predefined species defaults in FHI-aims (defaults2020). Nodes: Number of nodes with 72 cores per node and 256 GB memory. Inst.: Number of created instances of the auto-tuning mechanism (cf. Eq.~\eqref{eq:n_instances}). Init: Initialization time for the Fock exchange computation (only once per SCF cycle). Fock: Average HSE06 Fock exchange computation time per SCF iteration. KS: Average time per SCF iteration for the Solution of the Kohn-Sham equations. M: Estimated maximum memory per node in GB. The references point to the publications from which the structural models were taken. The visualization of the structures are shown in the SI.}
\label{tab:bulk}
\end{table}

\begin{table}[ht]
\centering
\begin{tabular}{|l|r|r|r|r|r|r|r|r|}
         \hline
         \multicolumn{9}{|c|}{\textbf{Nanosystems}} \\ \hline
         Species & Nodes & \small{Inst.} & Init (s) & Fock (s) & KS (s) & n\_basis & n\_states & M (GB) \\ \hline
         \multicolumn{9}{|c|}{Carbon Nanotube: C$_{2,000}$ (2,000 atoms), 1$\times$1$\times$1 k-grid, Ref.~\onlinecite{kecceli2016}} \\ \hline
         \multirow{2}{*}{light} & 4 & 4 &   41.446 &   21.119 &   30.698 & \multirow{2}{*}{28,000} & \multirow{2}{*}{12,000} & 124.36 \\
         & 8 & 8 &   42.297 &   11.019 &   17.887 & & & 123.59   \\ \hline
         interm. & 8 & 1 &  122.634 &  194.599 &   42.215 & 60,000 & 12,000 & 168.70 \\ \hline
         \multicolumn{9}{|c|}{Carbon Nanowire: C$_{2,000}$ (2,000 atoms), 1$\times$1$\times$1 k-grid, Ref.~\onlinecite{kecceli2016}} \\ \hline
         light & 8 & 4 &   62.211 &   57.638 &   20.003 & 28,000 & 12,000 & 112.40 \\ \hline
         interm. & 16 & 1 &  154.032 &  549.161 &   29.219 & 60,000 & 12,000 & 165.05 \\
         
         \hline
         \hline
        \multicolumn{9}{|c|}{\textbf{Clusters and Molecules}} \\ \hline
        Species & Nodes & \small{Inst.} & Init (s) & Fock (s) & KS (s) & n\_basis & n\_states & M (GB) \\ \hline
        \multicolumn{9}{|c|}{Solvated DNA: C$_{209}$H$_{10,166}$N$_{88}$Na$_{20}$O$_{5,110}$P$_{20}$ (15,613 atoms), Ref.~\onlinecite{mohr2017}}\\ \hline
        \multirow{2}{*}{light} & 32 & 2 &  132.269 &  254.326 &  546.280 & \multirow{2}{*}{127,328} & \multirow{2}{*}{58,308} & 180.00 \\
        & 64 & 8 &  129.111 &  104.250 &  272.003 &  &  & 188.33 \\ \hline
        interm. & 128 & 1 &  359.336 &  736.282 &  320.969 & 275,236 & 58,308 & 210.79 \\ \hline
        \multicolumn{9}{|c|}{Water drop: H$_{1,200}$O$_{600}$ (1,800 atoms), Ref.~\onlinecite{mohr2017}} \\ \hline
        light & 2 & 4 &   43.654 &   14.111 &    9.484 & 14,400 & 6,600 & 107.65 \\ \hline
        interm. & 4 & 1 &   89.275 &   84.418 &   11.444 & 31,200 & 6,600 & 131.53 \\ \hline
        \multicolumn{9}{|c|}{Silicon Wire: H$_{246}$Si$_{460}$ (706 atoms), Ref.~\onlinecite{mohr2017}} \\ \hline
        light & 4 & 4 &   33.700 &   42.645 &    3.805 & 12,730 & 5,092 & 90.82 \\ \hline
        interm. & 4 & 2 &   80.333 &  134.623 &    4.814 & 18,346 & 5,092 & 126.88 \\ \hline
        \multicolumn{9}{|c|}{Ac-Lys-Ala19-H$^+$: C$_{65}$H$_{112}$N$_{21}$O$_{22}$ (220 atoms)} \\ \hline
        light & 2 & 16 &    6.567 &    0.862 &    0.123 & 2,072 & 904 & 35.25 \\ \hline
        interm. & 2 & 8 &   17.415 &   10.134 &    0.215 & 4,472 & 904 & 82.98 \\ \hline
\end{tabular}
\caption{All calculations are carried out with the HSE06($\alpha$=0,$\beta$=0.25,$\omega$=0.11 Bohr$^{-1}$) functional. Species: Predefined species defaults in FHI-aims (defaults2020). Nodes: Number of nodes with 72 cores per node and 256 GB memory. Inst.: Number of created instances of the auto-tuning mechanism (cf. Eq.~\eqref{eq:n_instances}). Init: Initialization time for the Fock exchange computation (only once per SCF cycle). Fock: Average HSE06 Fock exchange computation time per SCF iteration. KS: Average time per SCF iteration for the Solution of the Kohn-Sham equations. M: Estimated maximum memory per node in GB. The references point to the publications from which the structural models were taken. The visualization of the structures are shown in the SI.}
    \label{tab:nano_nonperiodic}
\end{table}

\subsection{Computation of Forces and Stress}
We showcase the runtimes for the evaluation of the forces and stress for a small subset of the above described benchmark systems, as listed in Table~\ref{tab:force_stress}. As mentioned above, the memory consumption for force and stress computations are considerably larger. We use the same number of nodes for energy, force, and force+stress evaluation to allow for a unbiased comparison of runtimes. For Fe$_2$O$_3$, we find that the computation of the force and stress components increases the memory consumption significantly. However, the difference between a forces-only and a force+stress computation is small. This is due to the fact that the stress components are not computed concurrently, but in serial batches, see Sec.~\ref{SSecForceStress}, which in turn results in higher runtimes. The number of SCF iterations also increases, since an additional SCF step is needed for the force evaluation and two additional steps are needed for the stress evaluation. Note that the stress evaluation takes a few additional SCF steps, since FHI-aims slightly tightens the electronic convergence criteria for the stress evaluation by default. 
Similar behavior can be observed for the water system. It appears that energy and force computations have similar memory consumption, but the additional memory usage of the energy-only computation is due to using more instances. Again, to avoid running out of memory, stress components are computed sequentially. 

\begin{table}[ht]
\centering
\begin{tabular}{|c|r|r|r|r|r|r|r|r|r|}
\hline
\multicolumn{10}{|c|}{\textbf{Bulk systems}} \\
\hline
Mode & \#N & Inst. & Init (s) & Fock (s) & KS (s) & Iter (s) & f/s (s) & \#SCF & M (GB) \\ \hline
\multicolumn{10}{|c|}{Hematite: Fe$_4$O$_{6}$ (10 atoms), 10$\times$10$\times$10 k-grid, intermediate, collinear spin} \\ \hline
-/- & 1 & 3 &   48.562 &   84.220 &    0.410 & 89.041 & - & 17 & 89.34  \\
\hline
f / - & 1 & 2 &  268.248 &   78.577 &    0.416 & 82.642 & 108.220 & 18 & 181.67 \\ \hline
f / s & 1 & 2 &  267.335 &   75.602 &    0.410 & 80.278 & 575.318 & 23 &  200.11 \\ \hline
\multicolumn{10}{|c|}{Water (Liquid): H$_{128}$O$_{64}$ (192 atoms), 4$\times$4$\times$4 k-grid} \\ \hline
- / - & 4 & 8 &   39.000 &   17.913 &    2.339 & 24.349 &  - & 11 & 172.76 \\
f / - & 4 & 2 &   65.442 &   28.589 &    2.645 & 35.610 & 61.176 & 12 & 184.58 \\ \hline
f / s & 4 & 2 &   67.330 &   28.217 &    2.655 & 35.783 & 278.821 & 15 & 229.16 \\ \hline
\hline
\multicolumn{10}{|c|}{\textbf{Molecules}} \\ \hline
\multicolumn{10}{|c|}{Ac-Lys-Ala19-H$^+$: C$_{65}$H$_{112}$N$_{21}$O$_{22}$ (220 atoms), intermediate} \\ \hline
-  & 2 & 8 &   17.415 &   10.134 &    0.214 & 15.619 & - & 12 & 82.21 \\ \hline
f  & 2 & 2 &   27.176 &   11.022 &    0.228 & 15.145 & 52.991 & 13 & 87.16 \\ \hline

\end{tabular}
\caption{Runtime comparison for the HSE06($\alpha$=0,$\beta$=0.25,$\omega$=0.11 Bohr$^{-1}$) force and stress evaluation using intermediate species defaults. Inst.: Number of created instances of the auto-tuning mechanism (cf. Eq.~\eqref{eq:n_instances}). Init: Initialization time for the Fock exchange computation (only once per SCF cycle). Fock: Average HSE06 Fock exchange computation time per SCF iteration. KS: Average time per SCF iteration for the Solution of the Kohn-Sham equations. Iter: Average time for a full SCF iteration. \#SCF: Number of SCF iterations until electronic convergence has been achieved. M: Estimated maximum memory per node in GB.}
    \label{tab:force_stress}
\end{table}

\subsection{Performance comparison of the  PBE, PBE0, and HSE06 functionals}

This section focuses on the runtime differences between the semi-local functional PBE and the hybrid density functionals HSE06 and PBE0. We select two bulk materials (hematite, 10 atoms, and liquid water, 192 atoms) and a molecule (lysine-teminated polyalanine helix, 220 atoms) to quantify the differences explicitly for each functional. We run the same system for the three functionals on the same number of nodes to allow for an easy runtime comparison, even though PBE and HSE06 would actually need far fewer computational resources than PBE0. 

The run-time differences can be significant for dense materials with heavier elements and few atoms - especially between the hybrid density functionals PBE0 and HSE06. The timing for the real-space evaluation of the EXX contribution is directly related to the extent of its Coulomb kernel. The PBE0 functional uses a bare Coulomb potential and, thus, its extent is formally infinite, see Sec.~\ref{SecIntro}. In practice, the extent is still limited by the sparsity of the density matrix and by the overlap of the basis functions due to the finite extent of the employed atom-centered orbitals. However, the overlap for the relevant basis pairs in the 4-center-2-electron integrals extends to several layers of nearest neighbors. In contrast, the HSE06 functional uses a screened Coulomb potential leading to only a short-range EXX contribution, see  Sec.~\ref{SecIntro}. The "standard" screening parameter for HSE06 is 0.2$\mathring{A}^{-1}$ as given in Ref.~\onlinecite{Krukau2006}. In turn, the exchange contribution is limited  to pairs of closest neighbors in most practical cases. Thus, the sparsity of all matrices increases and leads to smaller workloads and hence lower runtimes. As shown in Tab.~\ref{tab:PBE_PBE0_HSE06}, the difference between PBE0 and HSE06 is most pronounced for dense materials -- in our case the Hematite crystal. HSE06 is a factor of six faster compared to PBE0. In contrast, for a small molecule, the runtime difference between HSE06 and PBE0 is negligible. 

The semi-local GGA functional PBE is faster by a factor of 18 compared to HSE06 for the Hematite crystal unit cell. The differences between HSE06 and PBE decrease for large systems and lighter elements, as well as for less dense materials. This can be observed for both the periodic water system and the molecule Ac-Lys-Ala19-H+. As the system size increases, the direct eigensolver ELPA will eventually dominate the runtimes, while, for smaller systems, the HSE06 exchange evaluation will dominate the runtime. For the largest systems listed in Tables~\ref{tab:bulk} and \ref{tab:nano_nonperiodic}, the PBE runtime can be directly estimated. For those calculations, the eigensolver and the HSE06 exchange evaluation are by far the dominant steps in the calculation, so the total HSE06 runtime per SCF iteration is approximately the sum of both. In turn, the PBE runtime can be estimated by using only the timings for the solution of the KS equations. For example, the $4\times4\times4$ paracetamol supercell is roughly only a factor of 1.5 faster when using PBE compared to the HSE06 calculation. 

\begin{table}[ht]
\centering
\begin{tabular}{|l|r|r|r|r|r|r|r|r|}
\hline
\multicolumn{9}{|c|}{\textbf{Bulk systems}} \\
\hline
Functional & Nodes & Inst. & Init (s) & Fock (s) & KS (s) & Iter (s) & \#SCF & M (GB) \\ \hline
\multicolumn{9}{|c|}{Hematite: Fe$_4$O$_{6}$ (10 atoms), 10$\times$10$\times$10 k-grid, intermediate, collinear spin} \\ \hline
PBE & 2 & - &    - &      - &    0.191 & 2.495 & 15 & 21.27 \\ \hline
HSE06 & 2 & 6 &   27.476 &   40.527 &    0.227 & 43.497 & 17 & 96.19 \\ \hline
PBE0 & 2 & 2 &   102.815 &  241.563 &    0.238 & 243.340 & 18 & 220.02 \\ \hline

\multicolumn{9}{|c|}{Water (Liquid): H$_{128}$O$_{64}$ (192 atoms), 4$\times$4$\times$4 k-grid} \\ \hline
PBE & 5 & - &    - &      - &    1.092 & 5.515 &  12 & 21.51 \\ \hline
HSE06 & 5 & 10 &   32.628 &   14.824 &    1.209 & 19.360 & 11 & 157.55 \\ \hline
PBE0 & 5 & 1 &   34.425 &   40.576 &    1.201 & 45.078 & 11 & 156.77 \\ \hline
\hline
\multicolumn{9}{|c|}{\textbf{Molecules}} \\ \hline
\multicolumn{9}{|c|}{Ac-Lys-Ala19-H$^+$: C$_{65}$H$_{112}$N$_{21}$O$_{22}$ (220 atoms), intermediate} \\ \hline
PBE & 2 & - &    - &      - &    0.218 & 5.372 & 13 & 21.82 \\ \hline
HSE06 & 2 & 8 &   17.415 &   10.134 &    0.215 & 15.619 & 12 & 82.21 \\ \hline
PBE0 & 2 & 8 &   16.505 &   12.803 &    0.228 & 18.191 & 12 & 75.52 \\ \hline

\end{tabular}
\caption{Runtime comparison for the PBE, PBE0($\alpha$=0.25,$\beta$=0.0), and HSE06($\alpha$=0,$\beta$=0.25,$\omega$=0.11 Bohr$^{-1}$) functional for the intermediate species defaults. Inst.: Number of created instances of the auto-tuning mechanism (cf. Eq.~\eqref{eq:n_instances}). Init: Initialization time for the Fock exchange computation (only once per SCF cycle). Fock: Average HSE06 Fock exchange computation time per SCF iteration. KS: Average time per SCF iteration for the Solution of the Kohn-Sham equations. Iter: Average time for a single SCF iteration. \#SCF: Number of SCF iterations until electronic convergence has been achieved. M: Estimated maximum memory per node in GB.}
    \label{tab:PBE_PBE0_HSE06}
\end{table}

\subsection{Limiting factors for the weak scaling behavior} 
\begin{figure}
    \centering
    \includegraphics[width=0.47\textwidth]{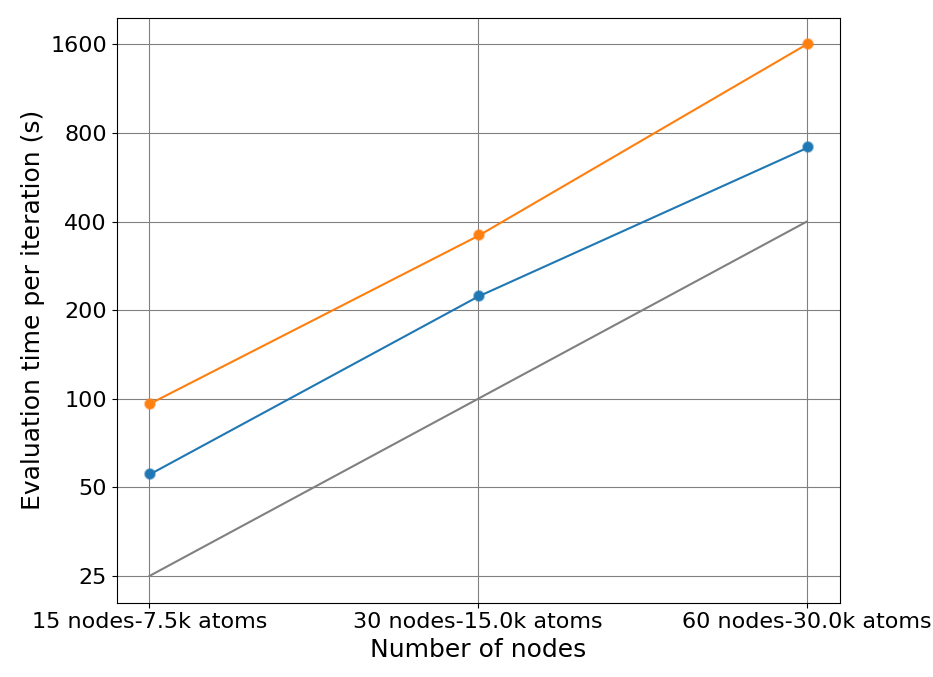}
    \caption{Weak scaling behavior for the  HSE06 Fock exchange evaluation (blue line) and the solution of the KS equations (orange line) for extreme supercells of the  Ice XI with 7.5k 15k, 30k atoms. The graph shows the timings per SCF iteration in seconds in a double logarithmic plot. The horizontal grey lines indicate ideal weak scaling and the grey diagonal line shows $O(N^2)$ scaling. The calculations were run on nodes with 256 GB RAM (first two data points from left to right), and with 512 GB RAM (last data point), respectively.}
    \label{fig:scaling_limits}
\end{figure}

The dense storage of the Hamiltonian, overlap, and Fock-type matrices, including the copies needed for the Pulay mixing, grow with $O(n_\text{basis}^2)$. Dense storage of these matrices is needed since our examples use a dense eigenvalue solver~(ELPA). Therefore these matrices become the memory bottleneck for large-scale simulations. For the exchange matrix evaluation, we observe the $O(N^2)$ behavior for our largest calculations,~i.e.,~for the Ice XI supercells with up to 30,000 atoms, as shown in Fig.~\ref{fig:scaling_limits}. Note that it is well known that all of the matrices involved become sparse for very large systems with a gap, in particular the density matrix and the closely related exchange matrix. However, this sparsity has not yet been exploited in the current implementation and will be part of future work. 

\section{Conclusions}

For the localized resolution of identity approach to hybrid density functional approximations, RI-LVL~\cite{ren2012,ihrig2015}, as implemented in the FHI-aims code, a thorough analysis of memory requirements and code efficiency was performed. An improved distributed storage algorithm was implemented using features of the MPI-3 standard. This allows us to exploit shared memory access on individual nodes for the storage of arrays that are common to each MPI task, while remaining entirely within the MPI paradigm. Other code parts were refactored accordingly, including a more sophisticated load balancing approach that relies on reducing communication by using the available memory to spawn independent ``instances'' (subgroups of MPI tasks) across which the exchange operator is evaluated. As a consequence, a drastic reduction with respect to memory requirements and a massive increase in code performance was achieved. For instance, the required memory per node now shows an almost optimal inverse scaling with respect to the number of employed nodes. These improvements are shown to enable the handling of system sizes $\gtrsim$ 10,000 atoms (more than 30,000 atoms in the largest example considered), extending the reach of hybrid density functional theory on standard CPU-based hardware far beyond the limits of what was previously possible, to our knowledge, in any electronic structure code. Furthermore, the improved distributed storage leads to better load-balance and reduced communication pressure. In turn, this results in an almost perfect, linear scaling of the computational effort with respect to system size and in a virtually ideal speedup with respect to the node count. For large system sizes, these improvements lead to a reduction of the runtimes by two orders of magnitude and more compared to the previous, already optimized implementation in FHI-aims. Regarding the limits the current implementation, we observe $O(N^2)$ weak scaling behavior for system sizes $\gtrsim$ 10,000 atoms. Nevertheless, the almost ideal strong scaling behavior is still present even in this regime of extremely large system sizes. The methods and algorithms presented are general and could be implemented in any electronic structure code that relies on localized basis functions. 

\section*{Code and data availability}

The FHI-aims code is an academic community code and available to any academic group, including its source code, for a voluntary license fee, enabling, access to the full sources and development thereof by any academic research group. All data that supports this work is openly available from the NOMAD data base. The corresponding URLs are listed in the SI.

\section*{Acknowledgements}

This work was funded by the NOMAD Center of Excellence (European Union’s Horizon 2020 research and innovation program, Grant Agreement No. 951786) and by the ERC Advanced Grant TEC1p (European Research Council, Grant Agreement No. 740233).


%
%

%


\bibliography{references}

\end{document}


\title{Supplemental Information}

\author{Sebastian Kokott}
\affiliation{The NOMAD Laboratory at the Fritz Haber Institute of the Max-Planck-Gesellschaft and IRIS Adlershof of the Humboldt-Universit\"{a}t zu Berlin, Germany}

\author{Florian Merz}
\affiliation{Lenovo HPC Innovation Center, Stuttgart, Germany}

\author{Yi Yao}
\affiliation{Thomas Lord Department of Mechanical Engineering and Material Science, Duke University, Durham, North Carolina 27708, USA}

\author{Christian Carbogno}
\affiliation{The NOMAD Laboratory at the Fritz Haber Institute of the Max-Planck-Gesellschaft and IRIS Adlershof of the Humboldt-Universit\"{a}t zu Berlin, Germany}

\author{Mariana Rossi}
\affiliation{MPI for the Structure and Dynamics of Matter, Luruper Chaussee 149, 22761 Hamburg, Germany}

\author{Ville Havu}
\affiliation{Department of Applied Physics, School of Science, Aalto University, Espoo, Finland}

\author{Markus Rampp}
\affiliation{Max Planck Computing and Data Facility, 85748 Garching, Germany}

\author{Matthias Scheffler}
\affiliation{The NOMAD Laboratory at the Fritz Haber Institute of the Max-Planck-Gesellschaft and IRIS Adlershof of the Humboldt-Universit\"{a}t zu Berlin, Germany}

\author{Volker Blum}
\affiliation{Thomas Lord Department of Mechanical Engineering and Material Science, Duke University, Durham, North Carolina 27708, USA}
\affiliation{Department of Chemistry, Duke University, Durham, North Carolina 27708, USA}

\date{\today}

\maketitle

\section{Benchmark Set}

\begin{figure}
    \centering
    \includegraphics[width=0.8\textwidth]{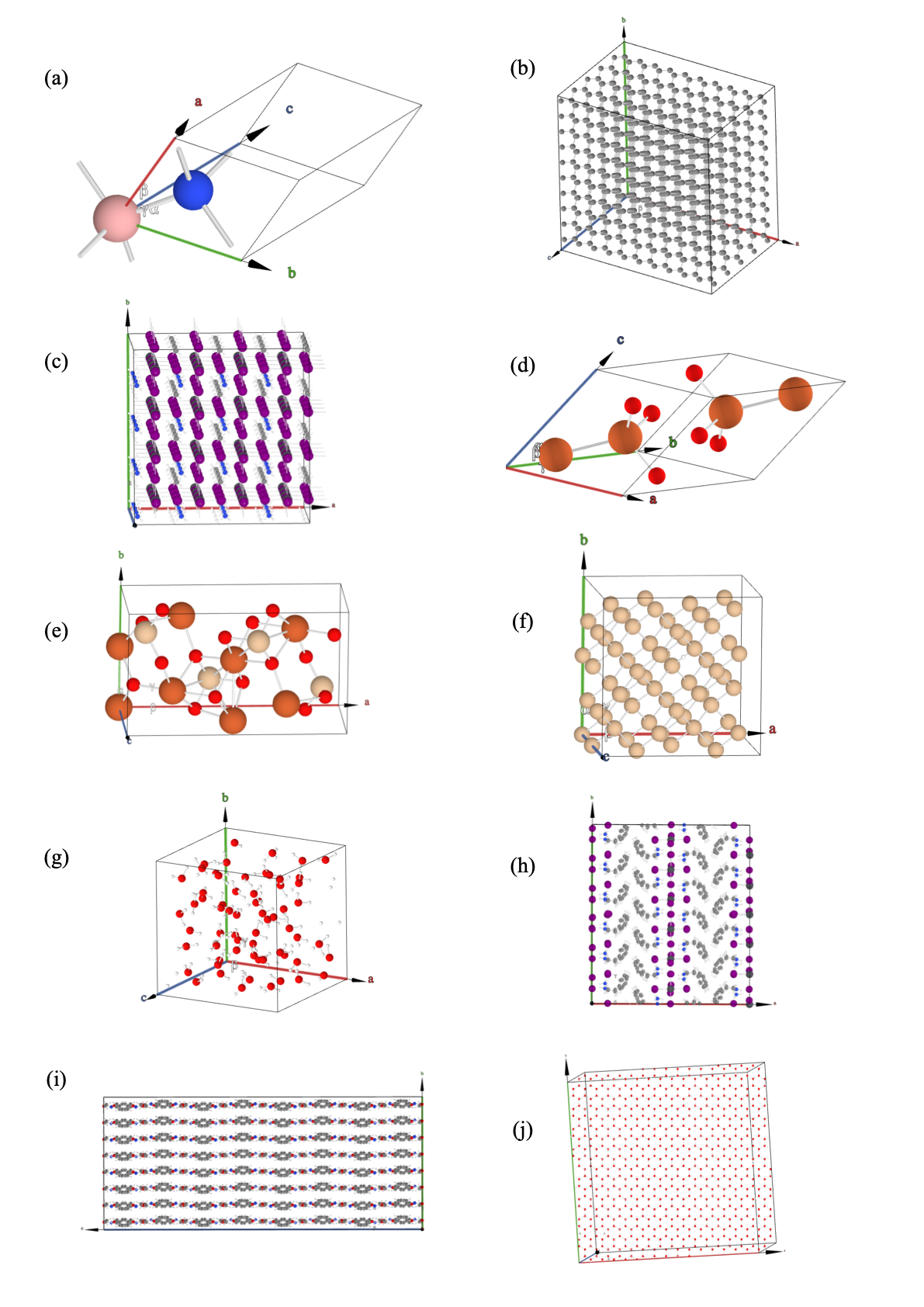}
    \caption{Visualization of the bulk systems. (a) Boron Nitride (2 atoms) (b) carbon diamond supercell (2,000 atoms). (c) MAPI supercell. (d) Hematite unit cell (10 atoms). (e) Fayalite unit cell (28 atoms). (f) Silicon supercell with Hydrogen interstitial (65 atoms). (g) Liquide water (192 atoms). (h) PEPI supercell with a defect complex containing a lead vacancy and two Bi substitutional defects on lead sides nearby (3,383 atoms). (i) Paracetamol super cell (10,240 atoms). (j) Ice XI supercell (30,576 atoms). All figures are created using the graphical user interface GIMS.} 
    \label{fig:bulk}
\end{figure}

\begin{figure}
    \centering
    \includegraphics[width=0.8\textwidth]{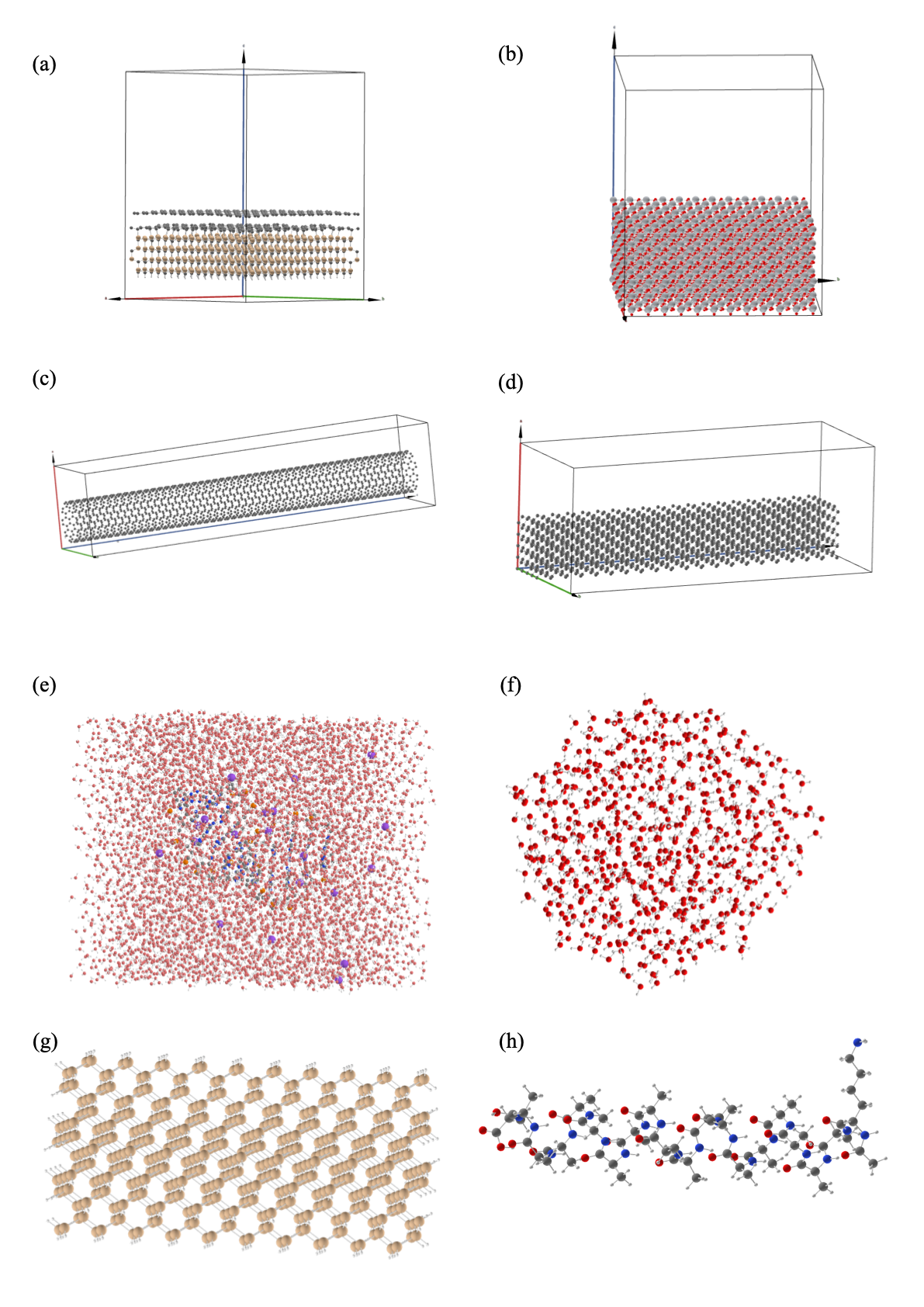}
    \caption{Visualization of the bulk systems. (a) Hydrogen terminated Silicon Carbide Slab with a graphene sheet on top (1,648 atoms). (b) TiO$_2$ slab (100) slab model (3,456 atoms). (c) Carbon nanotube (2,000 atoms). (d) Carbon nanowire (2,000 atoms). (e) DNA part solvated in saline water (15,613 atoms). (f) Water drop (1,800 atoms). (g) Silicon wire passivated with Hyrdogen (706 atoms). (h) Ac-Lys-Ala19-H$^+$ molecule. All figures are created using the graphical user interface GIMS.} 
    \label{fig:surf_nano_molec}
\end{figure}

The benchmark set used in this work is mainly based on a subset of structures from the \href{https://gitlab.com/elsi_project/elsi_benchmark}{ELSI benchmark repository} extended by the ``PEPI with Defect'' structure, the paracetamol supercell and the ICE XI supercell structure.  The table \ref{tab:references} lists the reference to the original publication of the used structures. Each table from the main text has a corresponding table here in the SI that lists the NOMAD entry IDs and URLs, from which the full data can be freely downloaded. In this SI, this are the tables \ref{tab:bulk}, \ref{tab:bulk+surface}, \ref{tab:nano_nonperiodic}, \ref{tab:force_stress}, and \ref{tab:PBE_PBE0_HSE06}. A visualization of benchmark systems can be found in Figs.~\ref{fig:bulk} and \ref{fig:surf_nano_molec}.

In the following subsections, we list and specify further details for the generation of the input structures. The geometry specifications of all structures is according to the FHI-aims input format \texttt{geometry.in}.

\begin{table}[]
    \centering
    \begin{tabular}{l|l}
         System Name & Reference \\ \hline \hline
         Boron Nitride &  Pearson's Handbook (1985) \\ \hline
         Carbon Diamond supercell & ACM TOMS Volume 33, No. 2, Article 9 (2007) \\ 
          &J. Comput. Chem. 37, 448 (2016) \\ \hline
         Cubic MAPI supercell & Phys Chem Chem Phys 17, 31360 (2015) \\ \hline
         Hematite Fe4O6 & springer materials ID sd\_0314193  \\ \hline
         Fayalite & springer materials ID sd\_0375064 \\ \hline
         Hydrogen interstitial in Silicon & J. Phys. Condens. Matter 26, 305503 (2014) \\ \hline
         Water (Liquid) &  J. Phys. Condens. Matter 26, 305503 (2014) \\ \hline
         PEPI with Defect & P R X Energy 2 (2023): 023010. \\ \hline
         Paracetamol supercell (Form II) & Derived from Phys. Rev. Lett. 117, 115702 \\ \hline
         Ice XI supercells & Derived from materials project ID mp-697111 \\ \hline
         Graphene on SiC & Phys Rev Lett 111, 065502 (2013) \\ \hline
         TiO2 slab & \href{https://gitlab.com/elsi_project/elsi_benchmark/-/tree/master/Periodic/Surfaces/TiO2?ref_type=heads}{ELSI benchmark: TiO2 slab}\\ \hline
         Carbon Nanotube and & ACM TOMS Volume 33, No. 2, Article 9 (2007) \\ 
         Carbon Nanowire&J. Comput. Chem. 37, 448 (2016) \\ \hline
         Solvated DNA, Water drop, & J. Chem. Theory Comput. 2017, 13, 10, 4684–4698 \\  
         and Silcon wire  & \\\hline
         Ac-Lys-Ala19-H+ & \href{https://gitlab.com/elsi_project/elsi_benchmark/-/tree/master/NonPeriodic/IsolatedMolecules/Ac-Lys-Ala19-H?ref_type=heads}{ELSI benchmark: Ac-Lys-Ala19-H}
         
    \end{tabular}
    \caption{List of original references for the systems that have been used in the benchmark.}
    \label{tab:references}
\end{table}

\begin{table}[ht]
\begin{tabular}{|l|r|l|}

    \hline
    \multicolumn{3}{|c|}{\textbf{Bulk systems}} \\
    \hline
    Species & \#Nodes & NOMAD entry ID and URL \\ \hline
     \multicolumn{3}{|c|}{Boron Nitride: BN (2 atoms), 19$\times$19$\times$19 k-grid} \\ \hline \hline
     light & 1 &\href{https://nomad-lab.eu/prod/v1/gui/user/datasets/dataset/id/q9sEh4EpTSiTZlbW2KjgOw/entry/id/E8ZWi2VhEhyMWuOSYijo4CZxlCLZ}{E8ZWi2VhEhyMWuOSYijo4CZxlCLZ}  \\ \hline
     interm. & 1 & \href{https://nomad-lab.eu/prod/v1/gui/user/datasets/dataset/id/q9sEh4EpTSiTZlbW2KjgOw/entry/id/9AkB_FjX44pSXpUbCL89Z81QahXb}{9AkB\_FjX44pSXpUbCL89Z81QahXb} \\ \hline
     tight & 1 &\href{https://nomad-lab.eu/prod/v1/gui/user/datasets/dataset/id/q9sEh4EpTSiTZlbW2KjgOw/entry/id/kAMgp9GCLthR5gtBu9hrjhClr13p}{kAMgp9GCLthR5gtBu9hrjhClr13p} \\ 
     \hline
     \multicolumn{3}{|c|}{Carbon Diamond: C$_{2000}$ (2,000 atoms), 2$\times$2$\times$2 k-grid} \\ \hline \hline
     \multirow{2}{*}{light} &  16 & \href{https://nomad-lab.eu/prod/v1/gui/user/datasets/dataset/id/q9sEh4EpTSiTZlbW2KjgOw/entry/id/O02swjtjLHXxXZswcOLGXrhwk7Zo}{O02swjtjLHXxXZswcOLGXrhwk7Zo} \\
      & 32 & \href{https://nomad-lab.eu/prod/v1/gui/user/datasets/dataset/id/q9sEh4EpTSiTZlbW2KjgOw/entry/id/JOTdhckn-CGciLtk9YxbJk_2bu4p}{JOTdhckn-CGciLtk9YxbJk\_2bu4p} \\
     \hline
     interm. & 64 & \href{https://nomad-lab.eu/prod/v1/gui/user/datasets/dataset/id/q9sEh4EpTSiTZlbW2KjgOw/entry/id/oJC7UrcxRs8ff4gRYguDgAZ4nXci}{oJC7UrcxRs8ff4gRYguDgAZ4nXci}\\ 
     \hline
     \multicolumn{3}{|c|}{Cubic MAPI: C$_{64}$N$_{64}$H$_{384}$Pb$_{64}$I$_{192}$ (768 atoms), 2$\times$2$\times$2 k-grid} \\ \hline \hline
     light & 8 & \href{https://nomad-lab.eu/prod/v1/gui/user/datasets/dataset/id/q9sEh4EpTSiTZlbW2KjgOw/entry/id/VUCVFXh4ve42hCQgWwhg9v2vsQDK}{VUCVFXh4ve42hCQgWwhg9v2vsQDK} \\ \hline
     \multirow{2}{*}{interm.} & 8 & \href{https://nomad-lab.eu/prod/v1/gui/user/datasets/dataset/id/q9sEh4EpTSiTZlbW2KjgOw/entry/id/Y1LROCbYx-k1_-wQc_vQGSbWtwsY}{Y1LROCbYx-k1\_-wQc\_vQGSbWtwsY} \\
     & 16 & \href{https://nomad-lab.eu/prod/v1/gui/user/datasets/dataset/id/q9sEh4EpTSiTZlbW2KjgOw/entry/id/4REgAbbjV3_gRlHZADywcm5sciL9}{4REgAbbjV3\_gRlHZADywcm5sciL9} \\ \hline
     \multicolumn{3}{|c|}{Hematite: Fe$_4$O$_{6}$ (10 atoms), 10$\times$10$\times$10 k-grid, collinear spin} \\ \hline \hline
     light & 1 & \href{https://nomad-lab.eu/prod/v1/gui/user/datasets/dataset/id/q9sEh4EpTSiTZlbW2KjgOw/entry/id/p24RxGuXibfUaNSTSUUlR55ATtx3}{p24RxGuXibfUaNSTSUUlR55ATtx3} \\ \hline
     interm. & 1 & \href{https://nomad-lab.eu/prod/v1/gui/user/datasets/dataset/id/q9sEh4EpTSiTZlbW2KjgOw/entry/id/ZospAItO1pCzdo6RMMf5JOLNgN3W}{ZospAItO1pCzdo6RMMf5JOLNgN3W} \\ \hline
     \multicolumn{3}{|c|}{Fayalite: Fe$_8$Si$_4$O$_{16}$ (28 atoms), 4$\times$7$\times$8 k-grid, collinear spin} \\ \hline \hline
     light & 2 & \href{https://nomad-lab.eu/prod/v1/gui/user/datasets/dataset/id/q9sEh4EpTSiTZlbW2KjgOw/entry/id/TD4Q7LruSlbccXSmCgFwDNpaa0Yy}{TD4Q7LruSlbccXSmCgFwDNpaa0Yy} \\ \hline
     interm. & 2 & \href{https://nomad-lab.eu/prod/v1/gui/user/datasets/dataset/id/q9sEh4EpTSiTZlbW2KjgOw/entry/id/PQtN38R4NZQGg92iFv5-yBkHIKuw}{PQtN38R4NZQGg92iFv5-yBkHIKuw} \\ \hline
     \multicolumn{3}{|c|}{Hydrogen interstitial in Silicon: Si$_{64}$H$_1$ (65 atoms), 4$\times$4$\times$4 k-grid} \\ \hline \hline
     light & 1 & \href{https://nomad-lab.eu/prod/v1/gui/user/datasets/dataset/id/q9sEh4EpTSiTZlbW2KjgOw/entry/id/GavULI_vhfV7R7gbuwxr5D6rynIo}{GavULI\_vhfV7R7gbuwxr5D6rynIo} \\ \hline
     \multirow{2}{*}{interm.} & 1 & \href{https://nomad-lab.eu/prod/v1/gui/user/datasets/dataset/id/q9sEh4EpTSiTZlbW2KjgOw/entry/id/TB4r0BlD2r-Rfm8zl3b9p5IFFMLO}{TB4r0BlD2r-Rfm8zl3b9p5IFFMLO} \\
     & 2 & \href{https://nomad-lab.eu/prod/v1/gui/user/datasets/dataset/id/q9sEh4EpTSiTZlbW2KjgOw/entry/id/BlnAq8gPwnNBnnLkGc89Nvh2_wMM}{BlnAq8gPwnNBnnLkGc89Nvh2\_wMM} \\ \hline
     \multicolumn{3}{|c|}{Water (Liquid): H$_{128}$O$_{64}$ (192 atoms), 4$\times$4$\times$4 k-grid} \\ \hline \hline
     light & 1 & \href{https://nomad-lab.eu/prod/v1/gui/user/datasets/dataset/id/q9sEh4EpTSiTZlbW2KjgOw/entry/id/q7emeI6YfyTIZV8CcIukbvqjCsvH}{q7emeI6YfyTIZV8CcIukbvqjCsvH} \\ \hline
     interm. & 1 & \href{https://nomad-lab.eu/prod/v1/gui/user/datasets/dataset/id/q9sEh4EpTSiTZlbW2KjgOw/entry/id/iOQ6K1fBr0jWDRVxS4chaj_vlVbE}{iOQ6K1fBr0jWDRVxS4chaj\_vlVbE} \\ 
     \hline

\end{tabular}
\caption{List of NOMAD IDs and URLs for the bulk materials in the benchmark.}
\label{tab:bulk}
\end{table}

\begin{table}[b]
\begin{tabular}{|l|r|l|}

    \hline
    \multicolumn{3}{|c|}{\textbf{Bulk systems}} \\
    \hline
    Species & \#Nodes & NOMAD entry ID and URL \\ \hline
     \multicolumn{3}{|c|}{PEPI with Defect: C$_{1,152}$H$_{1,728}$Bi$_2$I$_{288}$N$_{144}$Pb$_{69}$ (3,383 atoms), 1$\times$1$\times$1 k-grid} \\ \hline
     \multirow{2}{*}{interm.}& 16 & \href{https://nomad-lab.eu/prod/v1/gui/user/datasets/dataset/id/q9sEh4EpTSiTZlbW2KjgOw/entry/id/es8GkU1I5SKkw1UNjKJCt1VMqCr4}{es8GkU1I5SKkw1UNjKJCt1VMqCr4} \\
      & 32 & \href{https://nomad-lab.eu/prod/v1/gui/user/datasets/dataset/id/q9sEh4EpTSiTZlbW2KjgOw/entry/id/eohQ1TBVeFjDQPmboU5h8S4ZOleC}{eohQ1TBVeFjDQPmboU5h8S4ZOleC} \\
     \hline
     \multicolumn{3}{|c|}{4$\times$4$\times$4 Paracetamol: C$_{4,096}$H$_{4,608}$N$_{512}$O$_{1,024}$ (10,240 atoms), 1$\times$1$\times$1 k-grid} \\ \hline \hline
     \multirow{2}{*}{light} & 16 &  \href{https://nomad-lab.eu/prod/v1/gui/user/datasets/dataset/id/q9sEh4EpTSiTZlbW2KjgOw/entry/id/5Aj9RcF5gRNyMg5KCLpdVry3a-ZO}{5Aj9RcF5gRNyMg5KCLpdVry3a-ZO} \\
      & 32 &  \href{https://nomad-lab.eu/prod/v1/gui/user/datasets/dataset/id/q9sEh4EpTSiTZlbW2KjgOw/entry/id/2hrAL0R5ELl2GqItVK8xfwwem2k7}{2hrAL0R5ELl2GqItVK8xfwwem2k7} \\ \hline
     interm. & 84 & \href{https://nomad-lab.eu/prod/v1/gui/user/datasets/dataset/id/q9sEh4EpTSiTZlbW2KjgOw/entry/id/CXxWs89YfVFb7bRLpPv6zxz6sH5F}{4loxk6ZnThuNvOzcIxPbiw}\\ \hline \hline
     \multicolumn{3}{|c|}{$^\ast$Supercell of Ice XI: H$_{20,384}$O$_{10,192}$ (30,576 atoms), 1$\times$1$\times$1 k-grid} \\ \hline
     light & 60 & \href{https://nomad-lab.eu/prod/v1/gui/user/datasets/dataset/id/q9sEh4EpTSiTZlbW2KjgOw/entry/id/JO_HRN6NWLAj6b7PP4kft0sxt9Tc}{JO\_HRN6NWLAj6b7PP4kft0sxt9Tc} \\ \hline
     
     \multicolumn{3}{|c|}{\textbf{Surfaces}} \\ \hline
     \multicolumn{3}{|c|}{Graphene on SiC slab: C$_{1,108}$Si$_{432}$H$_{108}$ (1,648 atoms), 2$\times$2$\times$1 k-grid} \\ \hline \hline
     light & 8 & \href{https://nomad-lab.eu/prod/v1/gui/user/datasets/dataset/id/q9sEh4EpTSiTZlbW2KjgOw/entry/id/lOUvmxgJdtFf-m5OpMxEfjZOoSG_}{lOUvmxgJdtFf-m5OpMxEfjZOoSG\_} \\ \hline
     interm. & 32 & \href{https://nomad-lab.eu/prod/v1/gui/user/datasets/dataset/id/q9sEh4EpTSiTZlbW2KjgOw/entry/id/wXy0XLj_F7Xf-jutlSD-Etg-idjy}{wXy0XLj\_F7Xf-jutlSD-Etg-idjy}\\
     \hline \hline
     \multicolumn{3}{|c|}{TiO$_2$ slab: Ti$_{1,152}$O$_{2,304}$ (3,456 atoms), 1$\times$1$\times$1 k-grid} \\ \hline 
     light & 32 & \href{https://nomad-lab.eu/prod/v1/gui/user/datasets/dataset/id/q9sEh4EpTSiTZlbW2KjgOw/entry/id/dSzIbTsrawhETcRix7A2dxibB6ZZ}{dSzIbTsrawhETcRix7A2dxibB6ZZ} \\ \hline
     interm. & 48 & \href{https://nomad-lab.eu/prod/v1/gui/user/datasets/dataset/id/q9sEh4EpTSiTZlbW2KjgOw/entry/id/8Nl-eLhc3EbN8GCv2tpg8r1FHyps}{8Nl-eLhc3EbN8GCv2tpg8r1FHyps} \\
     \hline

\end{tabular}
\caption{List of NOMAD IDs and URLs for the bulk materials and surface slabs in the benchmark.}
\label{tab:bulk+surface}
\end{table}

\begin{table}[b]
\centering
\begin{tabular}{|l|r|l|}
         \hline
         \multicolumn{3}{|c|}{\textbf{Nanosystems}} \\ \hline
         Species & \#Nodes & NOMAD entry ID and URL \\ \hline
         \multicolumn{3}{|c|}{Carbon Nanotube: C$_{2,000}$ (2,000 atoms), 1$\times$1$\times$1 k-grid} \\ \hline
         \multirow{2}{*}{light} & 4 & \href{https://nomad-lab.eu/prod/v1/gui/user/datasets/dataset/id/q9sEh4EpTSiTZlbW2KjgOw/entry/id/akVwSsFdERvxUO5WWhcAxPdbWDBO}{akVwSsFdERvxUO5WWhcAxPdbWDBO} \\
         & 8 & \href{https://nomad-lab.eu/prod/v1/gui/user/datasets/dataset/id/q9sEh4EpTSiTZlbW2KjgOw/entry/id/TJw1NTVIFRFHhqMXalrqwY2kGsld}{TJw1NTVIFRFHhqMXalrqwY2kGsld}   \\ \hline
         interm. & 8 & \href{https://nomad-lab.eu/prod/v1/gui/user/datasets/dataset/id/q9sEh4EpTSiTZlbW2KjgOw/entry/id/kO4ds2zAgqaNS-Bpx-GEmk7rl_Ux}{kO4ds2zAgqaNS-Bpx-GEmk7rl\_Ux}\\ \hline
         \multicolumn{3}{|c|}{Carbon Nanowire: C$_{2,000}$ (2,000 atoms), 1$\times$1$\times$1 k-grid} \\ \hline
         light & 8 & \href{https://nomad-lab.eu/prod/v1/gui/user/datasets/dataset/id/q9sEh4EpTSiTZlbW2KjgOw/entry/id/mU6ISpXKdzK705cbHAqlUc1XijHl}{mU6ISpXKdzK705cbHAqlUc1XijHl} \\ \hline
         interm. & 16 & \href{https://nomad-lab.eu/prod/v1/gui/user/datasets/dataset/id/q9sEh4EpTSiTZlbW2KjgOw/entry/id/PP41Pt4M--1v4zu42AbNDb9sVUwv}{PP41Pt4M--1v4zu42AbNDb9sVUwv} \\
         
         \hline
         \hline
        \multicolumn{3}{|c|}{\textbf{Clusters and Molecules}} \\ \hline
        Species & \#Nodes & NOMAD entry ID and URL \\ \hline
        \multicolumn{3}{|c|}{Solvated DNA: C$_{209}$H$_{10,166}$N$_{88}$Na$_{20}$O$_{5,110}$P$_{20}$ (15,613 atoms)}\\ \hline
        \multirow{2}{*}{light} & 32 & \href{https://nomad-lab.eu/prod/v1/gui/user/datasets/dataset/id/q9sEh4EpTSiTZlbW2KjgOw/entry/id/nenjt7sHmtkJ87dOL6sKsd64vp1y}{nenjt7sHmtkJ87dOL6sKsd64vp1y} \\
        & 64 & \href{https://nomad-lab.eu/prod/v1/gui/user/datasets/dataset/id/q9sEh4EpTSiTZlbW2KjgOw/entry/id/LqWdxW3ovoJ3O_wOI6fKF1Vivd2H}{LqWdxW3ovoJ3O\_wOI6fKF1Vivd2H} \\ \hline
        interm. & 128 & \href{https://nomad-lab.eu/prod/v1/gui/upload/id/9hOeJajCTvWikIxjwWWm9Q}{G9C5-7Tu797emJSyUKg9rBhcvzsK} \\ \hline
        \multicolumn{3}{|c|}{Water drop: H$_{1,200}$O$_{600}$ (1,800 atoms)} \\ \hline
        light & 2 & \href{https://nomad-lab.eu/prod/v1/gui/user/datasets/dataset/id/q9sEh4EpTSiTZlbW2KjgOw/entry/id/S8wGNkpf1OBcMmNsSBtllfd87GWI}{S8wGNkpf1OBcMmNsSBtllfd87GWI} \\ \hline
        interm. & 4 & \href{https://nomad-lab.eu/prod/v1/gui/user/datasets/dataset/id/q9sEh4EpTSiTZlbW2KjgOw/entry/id/-_Dyqqz6SbHyS0MW_8Pyv_ZLLSxB}{-\_Dyqqz6SbHyS0MW\_8Pyv\_ZLLSxB} \\ \hline
        \multicolumn{3}{|c|}{Silicon Wire: H$_{246}$Si$_{460}$ (706 atoms)} \\ \hline
        light & 4 & \href{https://nomad-lab.eu/prod/v1/gui/user/datasets/dataset/id/q9sEh4EpTSiTZlbW2KjgOw/entry/id/oW5CvAxrykrIkj4Vg89bUAeaMEv6}{oW5CvAxrykrIkj4Vg89bUAeaMEv6} \\ \hline
        interm. & 4 & \href{https://nomad-lab.eu/prod/v1/gui/user/datasets/dataset/id/q9sEh4EpTSiTZlbW2KjgOw/entry/id/DFBTatOdAe65AcHH_N8zDqqjT8b8}{DFBTatOdAe65AcHH\_N8zDqqjT8b8} \\ \hline
        \multicolumn{3}{|c|}{Ac-Lys-Ala19-H$^+$: C$_{65}$H$_{112}$N$_{21}$O$_{22}$ (220 atoms)} \\ \hline
        light & 2 & \href{https://nomad-lab.eu/prod/v1/gui/user/datasets/dataset/id/q9sEh4EpTSiTZlbW2KjgOw/entry/id/_gnLa6Uo7RatprDWV1HfpbF6VP-0}{\_gnLa6Uo7RatprDWV1HfpbF6VP-0} \\ \hline
        interm. & 2 & \href{https://nomad-lab.eu/prod/v1/gui/user/datasets/dataset/id/q9sEh4EpTSiTZlbW2KjgOw/entry/id/rgwAxy6Jms0niy0JQgmbPyn2Y-oI}{rgwAxy6Jms0niy0JQgmbPyn2Y-oI} \\ \hline
\end{tabular}
\caption{List of NOMAD IDs and URLs for the nanosystems, clusters, and molecules in the benchmark.}
    \label{tab:nano_nonperiodic}
\end{table}

\begin{table}[b]
\centering
\begin{tabular}{|c|r|l|}
\hline
\multicolumn{3}{|c|}{\textbf{Bulk systems}} \\
\hline
Mode & \#Nodes & NOMAD entry ID and URL  \\ \hline
\multicolumn{3}{|c|}{Hematite: Fe$_4$O$_{6}$ (10 atoms), 10$\times$10$\times$10 k-grid, intermediate, collinear spin} \\ \hline
-/- & 1 & \href{https://nomad-lab.eu/prod/v1/gui/user/datasets/dataset/id/q9sEh4EpTSiTZlbW2KjgOw/entry/id/ZospAItO1pCzdo6RMMf5JOLNgN3W}{ZospAItO1pCzdo6RMMf5JOLNgN3W} \\
\hline
f / - & 1 & \href{https://nomad-lab.eu/prod/v1/gui/user/datasets/dataset/id/q9sEh4EpTSiTZlbW2KjgOw/entry/id/KSbkBYD_QnGn-XufePb1FcMZhZ5_}{KSbkBYD\_QnGn-XufePb1FcMZhZ5\_} \\ \hline
f / s & 1 & \href{https://nomad-lab.eu/prod/v1/gui/user/datasets/dataset/id/q9sEh4EpTSiTZlbW2KjgOw/entry/id/-nQ4UxT0HPvZbdHvYPZRRHRxloTt}{-nQ4UxT0HPvZbdHvYPZRRHRxloTt} \\ \hline
\multicolumn{3}{|c|}{Water (Liquid): H$_{128}$O$_{64}$ (192 atoms), 4$\times$4$\times$4 k-grid} \\ \hline
- / - & 4 & \href{https://nomad-lab.eu/prod/v1/gui/user/datasets/dataset/id/q9sEh4EpTSiTZlbW2KjgOw/entry/id/pw0aUhROe0vQaS_q-oMzG1eX3D4t}{pw0aUhROe0vQaS\_q-oMzG1eX3D4t} \\ \hline
f / - & 4 & \href{https://nomad-lab.eu/prod/v1/gui/user/datasets/dataset/id/q9sEh4EpTSiTZlbW2KjgOw/entry/id/dKklmSN7zXsWSeCG79pCpPv1oopd}{dKklmSN7zXsWSeCG79pCpPv1oopd} \\ \hline
f / s & 4 & \href{https://nomad-lab.eu/prod/v1/gui/user/datasets/dataset/id/q9sEh4EpTSiTZlbW2KjgOw/entry/id/Xju3ZHaXZh9K_XkFfwnet4yabpSJ}{Xju3ZHaXZh9K\_XkFfwnet4yabpSJ} \\ \hline
\hline
\multicolumn{3}{|c|}{\textbf{Molecules}} \\ \hline
\multicolumn{3}{|c|}{Ac-Lys-Ala19-H$^+$: C$_{65}$H$_{112}$N$_{21}$O$_{22}$ (220 atoms), intermediate} \\ \hline
-  & 2 & \href{https://nomad-lab.eu/prod/v1/gui/user/datasets/dataset/id/q9sEh4EpTSiTZlbW2KjgOw/entry/id/rgwAxy6Jms0niy0JQgmbPyn2Y-oI}{rgwAxy6Jms0niy0JQgmbPyn2Y-oI} \\ \hline
f  & 2 & \href{https://nomad-lab.eu/prod/v1/gui/user/datasets/dataset/id/q9sEh4EpTSiTZlbW2KjgOw/entry/id/FvxR9w73mf-Ae_UcEJgXuObM7-is}{FvxR9w73mf-Ae\_UcEJgXuObM7-is} \\ \hline 

\end{tabular}
\caption{List of NOMAD IDs and URLs for the energy, force, and stress computations.}
    \label{tab:force_stress}
\end{table}

\begin{table}[b]
\centering
\begin{tabular}{|l|r|l|}
\hline
\multicolumn{3}{|c|}{\textbf{Bulk systems}} \\
\hline
Functional & \#Nodes & NOMAD entry ID and URL \\ \hline
\multicolumn{3}{|c|}{Hematite: Fe$_4$O$_{6}$ (10 atoms), 10$\times$10$\times$10 k-grid, intermediate, collinear spin} \\ \hline
PBE & 2 & \href{https://nomad-lab.eu/prod/v1/gui/user/datasets/dataset/id/q9sEh4EpTSiTZlbW2KjgOw/entry/id/6qU6cehHRqcPPdxCmtdxTmnvCkTB}{6qU6cehHRqcPPdxCmtdxTmnvCkTB} \\ \hline
HSE06 & 2 & \href{https://nomad-lab.eu/prod/v1/gui/user/datasets/dataset/id/q9sEh4EpTSiTZlbW2KjgOw/entry/id/F9sgIAxq1PEUWsnTt2iI2vK1sUy0}{F9sgIAxq1PEUWsnTt2iI2vK1sUy0} \\ \hline
PBE0 & 2 & \href{https://nomad-lab.eu/prod/v1/gui/user/datasets/dataset/id/q9sEh4EpTSiTZlbW2KjgOw/entry/id/LVzpzkLofCadB2UWw3GmomCyF_EF}{LVzpzkLofCadB2UWw3GmomCyF\_EF} \\ \hline

\multicolumn{3}{|c|}{Water (Liquid): H$_{128}$O$_{64}$ (192 atoms), 4$\times$4$\times$4 k-grid} \\ \hline
PBE & 5 & \href{https://nomad-lab.eu/prod/v1/gui/user/datasets/dataset/id/q9sEh4EpTSiTZlbW2KjgOw/entry/id/dai52NIQ0FyhTQwp_UK18gIXiyRq}{dai52NIQ0FyhTQwp\_UK18gIXiyRq} \\ \hline
HSE06 & 5 & \href{https://nomad-lab.eu/prod/v1/gui/user/datasets/dataset/id/q9sEh4EpTSiTZlbW2KjgOw/entry/id/pmT7HqAqWuBi5n3kKgMkCeUHEIpb}{pmT7HqAqWuBi5n3kKgMkCeUHEIpb} \\ \hline
PBE0 & 5 & \href{https://nomad-lab.eu/prod/v1/gui/user/datasets/dataset/id/q9sEh4EpTSiTZlbW2KjgOw/entry/id/c02rKq25k40AweV7FK6kyVaHLEAg}{c02rKq25k40AweV7FK6kyVaHLEAg} \\ \hline
\hline
\multicolumn{3}{|c|}{\textbf{Molecules}} \\ \hline
\multicolumn{3}{|c|}{Ac-Lys-Ala19-H$^+$: C$_{65}$H$_{112}$N$_{21}$O$_{22}$ (220 atoms), intermediate} \\ \hline
PBE & 2 & \href{https://nomad-lab.eu/prod/v1/gui/user/datasets/dataset/id/q9sEh4EpTSiTZlbW2KjgOw/entry/id/_e0U2GKSZAtsg9UWlYwfKZiPH6zj}{\_e0U2GKSZAtsg9UWlYwfKZiPH6zj} \\ \hline
HSE06 & 2 & \href{https://nomad-lab.eu/prod/v1/gui/user/datasets/dataset/id/q9sEh4EpTSiTZlbW2KjgOw/entry/id/rgwAxy6Jms0niy0JQgmbPyn2Y-oI}{rgwAxy6Jms0niy0JQgmbPyn2Y-oI} \\ \hline
PBE0 & 2 & \href{https://nomad-lab.eu/prod/v1/gui/user/datasets/dataset/id/q9sEh4EpTSiTZlbW2KjgOw/entry/id/-afSX20PZLYTrPsL-GmCkr4o6Nbe}{-afSX20PZLYTrPsL-GmCkr4o6Nbe} \\ \hline

\end{tabular}
\caption{List of NOMAD IDs and URLs for the runtime comparison for the PBE, PBE0($\alpha$=0.25,$\beta$=0.0), and HSE06($\alpha$=0,$\beta$=0.25,$\omega$=0.11 Bohr$^{-1}$) functional for the intermediate species defaults.}
    \label{tab:PBE_PBE0_HSE06}
\end{table}

\subsection{Boron Nitride}

\begin{verbatim}
lattice_vector 0.0000000000000000 1.8077500000000000 1.8077500000000000
lattice_vector 1.8077500000000000 0.0000000000000000 1.8077500000000000
lattice_vector 1.8077500000000000 1.8077500000000000 0.0000000000000000

atom_frac 0.0000000000000000 0.0000000000000000 0.0000000000000000 B
atom_frac 0.2500000000000000 0.2500000000000000 0.2500000000000000 N
\end{verbatim}

\subsection{Carbon Diamond supercell}

The Carbon Diamond supercell containing 2000 atoms is based on the following unit cell:

\begin{verbatim}
lattice_vector 0.0000000000000000 1.7810000214081378 1.7810000214081378
lattice_vector 1.7810000214081378 0.0000000000000000 1.7810000214081378
lattice_vector 1.7810000214081378 1.7810000214081378 0.0000000000000000

atom_frac 0.0000000000000000 0.0000000000000000 0.0000000000000000 C
atom_frac 0.2500000000000000 0.2500000000000000 0.2500000000000000 C
\end{verbatim}

The supercell lattice $(a_\text{sc},b_\text{sc},c_\text{sc})$ is generated by multiplying with the primitive lattice $(a_\text{pc},b_\text{pc},c_\text{pc})$ with the supercell matrix $\mathbf{x}$:

\begin{align}
    (a_\text{pc},b_\text{pc},c_\text{pc})\cdot\mathbf{x} = (a_\text{sc},b_\text{sc},c_\text{sc})^\text{T}
\end{align}
with 
\begin{align}
\mathbf{x} = 
\begin{pmatrix}
-5\sqrt{2} & 5\sqrt{2} & 5\\
5\sqrt{2} & -5\sqrt{2} & 5\\
5\sqrt{2} & 5\sqrt{2} & -5 
\end{pmatrix}
\end{align}

\subsection{Cubic MAPI supercell}

The Cubic MAPI (methyl ammonium lead iodide perovskite) 4$\times$4$\times$4 supercell containing 768 atoms is based on the following unit cell:

\begin{verbatim}
lattice_vector 6.3424000625000000 0.0000000000000000 0.0000000000000000
lattice_vector 0.0000000000000000 6.3424000625000000 0.0000000000000000
lattice_vector 0.0000000000000000 0.0000000000000000 6.3424000625000000
atom_frac 0.4985070174686732 0.4984753283607610 0.4984391279480255 Pb
atom_frac 0.9947067051243741 0.4985585744497498 0.4985239897502917 I
atom_frac 0.4933691056089858 0.9946781165777344 0.4985067920019747 I
atom_frac 0.4934096948374265 0.4933045519706207 0.9946418980330606 I
atom_frac 0.9636066959707023 0.9685230268852983 0.9735016928627009 C
atom_frac 0.0964661473134566 0.1028459790177418 0.1092681953078861 N
atom_frac 0.8648932142712814 0.8687516830305602 0.0746980652562706 H
atom_frac 0.0669248970527235 0.8708916836401468 0.8748263002291492 H
atom_frac 0.8648877825735546 0.0707791241212009 0.8747967353171048 H
atom_frac 0.1950609531184868 0.0158756953610258 0.2080049998028644 H
atom_frac 0.0084353203397443 0.2005150264439032 0.2079821657812684 H
atom_frac 0.1950536837883363 0.2024919488039627 0.0233524765531140 H
\end{verbatim}

\subsection{Hematite Fe$_4$O$_6$}

The Hematite unit cell has an anti-ferromagnetic spin order. Spin and charge initialization for the Hematite structure are given by the following input file.

\begin{verbatim}
lattice_vector  2.9069590000000000  0.0000000000000000 4.5766670000000000
lattice_vector -1.4534790000000000  2.5175000000000000 4.5766670000000000
lattice_vector -1.4534790000000000 -2.5175000000000000 4.5766670000000000
atom_frac 0.1478299626138310 0.1478300134676661 0.1478300134676661 Fe
    initial_charge         3.000000
    initial_moment        -5.000000
atom_frac 0.3521698863021943 0.3521700074494164 0.3521700074494164 Fe
    initial_charge         3.000000
    initial_moment         5.000000
atom_frac 0.6478298115298562 0.6478300343847486 0.6478300343847484 Fe
    initial_charge         3.000000
    initial_moment         5.000000
atom_frac 0.8521697352182194 0.8521700283664989 0.8521700283664987 Fe
    initial_charge         3.000000
    initial_moment        -5.000000
atom_frac 0.5561799458228389 0.9438200295675879 0.2499999699846684 O
    initial_charge        -2.000000
atom_frac 0.9438199030931863 0.2500000509324142 0.5561799913494946 O
    initial_charge        -2.000000
atom_frac 0.2499999244580125 0.5561799508756218 0.9438200700414608 O
    initial_charge        -2.000000
atom_frac 0.4438197520092117 0.0561800122665770 0.7500000718494965 O
    initial_charge        -2.000000
atom_frac 0.0561797947388639 0.7499999909017510 0.4438200504846703 O
    initial_charge        -2.000000
atom_frac 0.7499997733740377 0.4438200909585432 0.0561799717927041 O
    initial_charge        -2.000000
\end{verbatim}

\subsection{Fayalite}

The Fayalite (Fe8Si4O16) unit cell contains 28 atoms and also is initialized in a anti-ferromagnetic spin order. For this structure, the different spin initialization of the iron atoms occupying different Wyckoff-position was in particular important to obtain SCF convergence.

\begin{verbatim}
lattice_vector 10.5170000000000000 0.0000000000000000 0.0000000000000000
lattice_vector 0.0000000000000000 6.0560000000000000 0.0000000000000000
lattice_vector 0.0000000000000000 0.0000000000000000 4.8220000000000001
atom_frac 0.2227000095084150 0.2500000000000000 0.5085000000000000 Fe
    initial_moment  3.77
    initial_charge  2.0
atom_frac 0.2772999904915850 0.7500000000000000 0.0085000000000000 Fe
    initial_moment -3.77
    initial_charge  2.0
atom_frac 0.7772999904915850 0.7500000000000000 0.4915000000000000 Fe
    initial_moment  3.77
    initial_charge  2.0
atom_frac 0.7227000095084150 0.2500000000000000 0.9914999999999999 Fe
    initial_moment -3.77
    initial_charge  2.0
atom_frac 0.0000000000000000 0.0000000000000000 0.0000000000000000 Fe
    initial_moment -3.76
    initial_charge  2.0
atom_frac 0.4999999999999999 0.5000000000000000 0.5000000000000000 Fe
    initial_moment  3.76
    initial_charge  2.0
atom_frac 0.4999999999999999 0.0000000000000000 0.5000000000000000 Fe
    initial_moment  3.76
    initial_charge  2.0
atom_frac 0.0000000000000000 0.5000000000000000 0.0000000000000000 Fe
    initial_moment -3.76
    initial_charge  2.0
atom_frac 0.4060000000000000 0.2500000000000000 0.0735999585234343 Si
    initial_charge  4.0
atom_frac 0.0940000000000000 0.7500000000000000 0.5735999585234343 Si
    initial_charge  4.0
atom_frac 0.5940000000000000 0.7500000000000000 0.9264000414765657 Si
    initial_charge  4.0
atom_frac 0.9060000000000000 0.2500000000000000 0.4264000414765657 Si
    initial_charge  4.0
atom_frac 0.3372999904915850 0.0331999669749009 0.2221999170468685 O
    initial_charge -2.0
atom_frac 0.1627000095084150 0.5331999669749009 0.7221999170468685 O
    initial_charge -2.0
atom_frac 0.1627000095084150 0.9668000330250991 0.7221999170468685 O
    initial_charge -2.0
atom_frac 0.6627000095084150 0.5331999669749009 0.7778000829531314 O
    initial_charge -2.0
atom_frac 0.6627000095084150 0.9668000330250991 0.7778000829531314 O
    initial_charge -2.0
atom_frac 0.8372999904915851 0.4668000330250991 0.2778000829531315 O
    initial_charge -2.0
atom_frac 0.8372999904915851 0.0331999669749009 0.2778000829531315 O
    initial_charge -2.0
atom_frac 0.3372999904915850 0.4668000330250991 0.2221999170468685 O
    initial_charge -2.0
atom_frac 0.0531000285252448 0.2500000000000000 0.2789000414765657 O
    initial_charge -2.0
atom_frac 0.4468999714747551 0.7500000000000000 0.7789000414765658 O
    initial_charge -2.0
atom_frac 0.9468999714747551 0.7500000000000000 0.7210999585234342 O
    initial_charge -2.0
atom_frac 0.5531000285252449 0.2500000000000000 0.2210999585234343 O
    initial_charge -2.0
atom_frac 0.4084000190168299 0.2500000000000000 0.7346999170468685 O
    initial_charge -2.0
atom_frac 0.0915999809831701 0.7500000000000000 0.2346999170468685 O
    initial_charge -2.0
atom_frac 0.5915999809831701 0.7500000000000000 0.2653000829531315 O
    initial_charge -2.0
atom_frac 0.9084000190168298 0.2500000000000000 0.7653000829531315 O
    initial_charge -2.0
\end{verbatim}

\subsection{Hydrogen interstitial in Silicon}

The Si supercell is based on the following primitive unit cell:
\begin{verbatim}
lattice_vector 0.0000000000000000 2.7150000000000000 2.7150000000000000
lattice_vector 2.7150000000000000 0.0000000000000000 2.7150000000000000
lattice_vector 2.7150000000000000 2.7150000000000000 0.0000000000000000
atom_frac 0.0000000000000000 0.0000000000000000 0.0000000000000000 Si
atom_frac 0.2500000000000000 0.2500000000000000 0.2500000000000000 Si
\end{verbatim}

using the supercell matrix:

\begin{align}
\mathbf{x} = 
\begin{pmatrix}
-2 & \phantom{-}2 & \phantom{-}2\\
\phantom{-}2 & -2 & \phantom{-}2\\
\phantom{-}2 & \phantom{-}2 & -2 
\end{pmatrix}
\end{align}

and placing a Hydrogen atom at (6.10875, 6.10875, 6.10875) \AA.

\subsection{Water (Liquid)}

The cell contains 64 water molecules. The input structure can be found here: [ADD NOMAD URL]. In total, there are 192 atoms. 

\subsection{PEPI with Defect}

The structure is based on a (6x6) phenylethylammonium lead iodide (PEA2PbI4, or short: PEPI) with defect complex taken from Lu et al. (P R X Energy 2 (2023): 023010.; also cf. Figure S1.8 therein). The defect complex consists of a lead vacancy and bismuth substitutions of two nearby lead atoms. In total, the structure has 3,383 atoms.

\subsection{4×4×4 Paracetamol supercell}

The 4×4×4 supercell of the paracetamol crystal is based on the metastable form II unit cell (orthorhombic) containing 8 paracetamol molecules. Supercells are simply generated by multiples of the lattice vectors from the unit cell. 

\subsection{Supercell of Ice XI}
The Ice XI supercell is based on the following primitive unit cell:

\begin{verbatim}
lattice_vector 2.1246016568898547 -3.6755729842109162 0.0000000000000000
lattice_vector 2.1246016568898547  3.6755729842109162 0.0000000000000000
lattice_vector 0.0000000000000000  0.0000000000000000 6.9422303803022247
atom_frac 0.1710121421845457 0.8289878578154540 0.2942537831864080 H
atom_frac 0.8289878578154543 0.1710121421845459 0.7942537831864080 H
atom_frac 0.0481949478915413 0.9518050521084587 0.4896036778631453 H
atom_frac 0.9518050521084587 0.0481949478915413 0.9896036778631454 H
atom_frac 0.5861284580080457 0.0284750971435118 0.5147720031996131 H
atom_frac 0.9715249028564882 0.4138715419919543 0.5147720031996131 H
atom_frac 0.4138715419919541 0.9715249028564881 0.0147720031996132 H
atom_frac 0.0284750971435118 0.5861284580080457 0.0147720031996132 H
atom_frac 0.1796291960199634 0.8203708039800366 0.4413296411621622 O
atom_frac 0.8203708039800366 0.1796291960199634 0.9413296411621622 O
atom_frac 0.8458673171536840 0.1541326828463160 0.5672469013890502 O
atom_frac 0.1541326828463159 0.8458673171536841 0.0672469013890502 O
\end{verbatim}
using the supercell matrix:

\begin{align}
\mathbf{x} = 
\begin{pmatrix}
\phantom{-}15 & 16 & -1\\
-10 & \phantom{ }8 & \phantom{-}0\\
\phantom{- }1 & \phantom{ }8 &  \phantom{-}9
\end{pmatrix}
\end{align}

\section{Strong and weak scaling for the $GaAs$ supercells}

All data points for the scaling plots for the GaAs supercell are available from the following link:
\url{https://nomad-lab.eu/prod/v1/gui/user/uploads/upload/id/iIewzg6ARwqKE6595ZKhkw}

\section{Weak scaling behavior for the Ice XI supercells}

\begin{itemize}
    \item 7.5k atoms: \href{https://nomad-lab.eu/prod/v1/gui/user/datasets/dataset/id/q9sEh4EpTSiTZlbW2KjgOw/entry/id/tqhm4yoe1JQYy_As3Fb4q4eyKt_K}{tqhm4yoe1JQYy\_As3Fb4q4eyKt\_K}
    \item 15k atoms: \href{https://nomad-lab.eu/prod/v1/gui/user/datasets/dataset/id/q9sEh4EpTSiTZlbW2KjgOw/entry/id/NMsFQklAAchZr-xOxeYJUTJ8hzJl}{NMsFQklAAchZr-xOxeYJUTJ8hzJl}
    \item 30k atoms: \href{https://nomad-lab.eu/prod/v1/gui/user/datasets/dataset/id/q9sEh4EpTSiTZlbW2KjgOw/entry/id/MHKIDmn6ao4DTOx8Sk1Z9jrIdaNs}{MHKIDmn6ao4DTOx8Sk1Z9jrIdaNs}
\end{itemize}

\section{Benchmark results for the largest periodic structure  (Fig.~1)}

In Fig.~1, only the data point for the 15,288-atoms Ice XI supercell is missing in the above tables. This data can be found here: 
\begin{itemize}
    \item without forces: \href{https://nomad-lab.eu/prod/v1/gui/user/uploads/upload/id/C0Lh72eqQTSZry-o1JKxCQ/entry/id/xKupmS4Kg_duP7EXGLMWfzAHwGCT}{xKupmS4Kg\_duP7EXGLMWfzAHwGCT}
    \item with forces: \href{https://nomad-lab.eu/prod/v1/gui/user/uploads/upload/id/C0Lh72eqQTSZry-o1JKxCQ/entry/id/N-hw7S_do23vl6cp3iC8hsWeaULc}{N-hw7S\_do23vl6cp3iC8hsWeaULc}
\end{itemize}

All other point of Fig.~1 are listed in Tab.~\ref{tab:bulk+surface}. 

